\documentclass[11pt,fleqn]{article}       
\usepackage{geometry}
\usepackage[T1]{fontenc}
\usepackage[utf8]{inputenc}
\usepackage{authblk}
\usepackage{xcolor}
\usepackage{mathrsfs}
\usepackage[font=small,labelfont=bf]{caption}
\usepackage{graphicx}
\usepackage{multirow}
\usepackage{amsmath,amssymb,amsfonts}
\usepackage{verbatim}
\usepackage{bm}
\usepackage{color}
\usepackage{ulem}
\usepackage{mathtools}
\usepackage{cite}
\usepackage{colortbl}
\usepackage{cancel}
\usepackage[shortlabels]{enumitem}
\definecolor{light-gray}{gray}{0.85}
\usepackage[pdftex,colorlinks,pdfpagelabels]{hyperref}
\hypersetup{
   bookmarksnumbered,
   citecolor={blue},
   linkcolor={blue},
   urlcolor={blue},
   filecolor={blue}
} 

\newcommand{\eVdist}{\kern-0.06em}
\newcommand{\AddrTexas}{%
\textit{Department of Physics, The University of Texas at Austin, Austin, 78712 TX, USA}
}
\newcommand{\AddrStockholm}{
\textit{Oskar Klein Center for Cosmoparticle Physics, University of Stockholm, 10691 Stockholm, Sweden}
}
\newcommand{\AddrNordita}{
\textit{Nordita, KTH Royal Institute of Technology and Stockholm University, 10691 Stockholm, Sweden}
}

\usepackage{fancyhdr}
 \usepackage[bottom]{footmisc}
\fancypagestyle{plain}{%
    \fancyhead[R]{UTWI-30-2023, NORDITA-2023-045}
    
}
\date{}
\title{The Gravitational Wave Spectrum of Chain Inflation}

\author[1,2,3]{Katherine Freese\thanks{ktfreese@utexas.edu}}
\author[2]{Aliki Litsa\thanks{aliki.litsa@fysik.su.se}}
\author[1,2]{Martin Wolfgang Winkler\thanks{martin.winkler@austin.utexas.edu}}
\affil[1]{\AddrTexas}
\affil[2]{\AddrStockholm}
\affil[3]{\AddrNordita}

\begin{document}
\maketitle
\vspace*{0mm}
\begin{abstract}
Chain inflation is an alternative to slow-roll inflation in which the inflaton tunnels along a large number of consecutive minima in its potential. In this work we perform the first comprehensive calculation of the gravitational wave spectrum of chain inflation. In contrast to slow-roll inflation the latter does not stem from quantum fluctuations of the gravitational field during inflation, but rather from the bubble collisions during the first-order phase transitions associated with vacuum tunneling.  
Our calculation is performed within an effective theory of chain inflation which builds on an expansion of the tunneling rate capturing most of the available model space. The effective theory can be seen as chain inflation's analogue of the slow-roll expansion in rolling models of inflation.
We show that chain inflation produces a very characteristic double-peak spectrum: a faint high-frequency peak associated with the gravitational radiation emitted during inflation, and a strong low-frequency peak associated with the graceful exit from chain inflation (marking the transition to the radiation-dominated epoch). 
There exist very exciting prospects to test the gravitational wave signal from chain inflation at the aLIGO-aVIRGO-KAGRA network, at LISA and /or at pulsar timing array experiments.
A particularly intriguing possibility we point out is that chain inflation could be the source of the stochastic gravitational wave background recently detected by NANOGrav, PPTA, EPTA and CPTA. 
We also show that the gravitational wave signal of chain inflation is often accompanied by running/ higher running of the scalar spectral index to be tested at future Cosmic Microwave Background experiments.

\end{abstract}
\clearpage

\section{Introduction}

Cosmic inflation~\cite{Guth:1980zm} is one of the corner stones of modern cosmology. The rapid expansion of space resolves the infamous horizon, flatness and monopole problems. At the same time, it generates the density perturbations in the primordial plasma which constitute the seeds for structure formation and for the temperature anisotropies we observe in the Cosmic Microwave Background (CMB)~\cite{Mukhanov:1981xt,Hawking:1982cz,Starobinsky:1982ee,Guth:1982ec,Bardeen:1983qw}. In its most studied realization, inflation occurs through a scalar field which is displaced from its minimum and slowly rolls down its potential~\cite{Linde:1981mu,Albrecht:1982wi}.

\begin{figure}[h!]
\begin{center}
  \vspace{-5mm}
  \includegraphics[width=11.5cm]{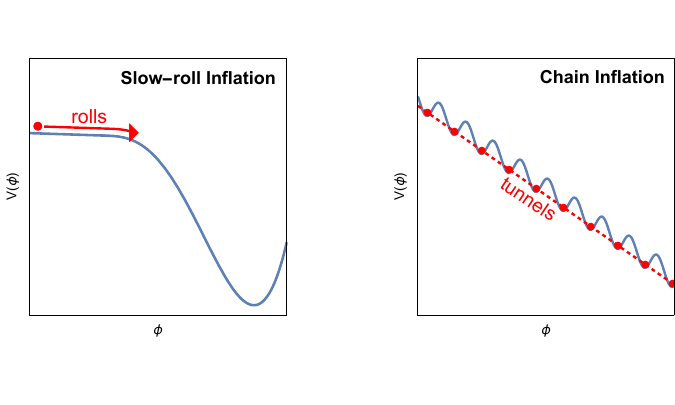}
\end{center}
\vspace{-1.2cm}
\caption{In slow-roll inflation (left panel) the inflaton rolls down its potential. In chain inflation it tunnels from minimum to minimum in its potential.} 
\label{fig:slowchain}
\end{figure}

However, there exists an alternative theory of the early Universe known as chain inflation~\cite{Freese:2004vs,Freese:2005kt,Ashoorioon:2008pj,Winkler:2020ape,Freese:2021noj} in which the inflaton -- rather than slowly rolling -- tunnels along a series of metastable minima in its potential (see Fig.~\ref{fig:slowchain}). While in slow-roll inflation, density perturbations originate from the quantum fluctuations of the inflaton~\cite{Mukhanov:1981xt}, in chain inflation they result from the probabilistic nature of tunneling~\cite{Feldstein:2006hm,Winkler:2020ape} -- the fact that different patches of the Universe undergo the vacuum transitions at slightly different times. Despite their differences, both inflation theories are equally compatible with all existing cosmological data. In particular, the observed nearly scale-invariant power spectrum of scalar perturbations can be realized in chain inflation through a tunneling rate which varies (at most) slowly from vacuum to vacuum along the chain~\cite{Winkler:2020ape,Freese:2021noj}. This suggests a quasi-periodic inflaton potential as it naturally occurs for axion fields\footnote{The term ``axion'' was originally introduced for the particle associated with the Peccei-Quinn solution to the strong-CP problem in the Standard Model~\cite{Peccei:1977hh}. However, in the more recent literature it is commonly employed more generally for pseudoscalar fields with an approximate shift symmetry. We follow this more general convention.} in extensions of the Standard Model of particle physics~\cite{Freese:2005kt,Ashoorioon:2008pj,Winkler:2020ape,Freese:2021noj}.

In previous work, we studied density perturbations arising in chain inflation~\cite{Winkler:2020ape,Freese:2021noj}.
We found an upper bound on the inflation scale where observables in the CMB are produced, $V^{1/4}_*\lesssim 10^{12}\:\text{GeV}$, in order to satisfy constraints on the scalar power spectrum, while finding consistent tunneling solutions.  Further, the reheating temperature after the last transition should be high enough to allow for successful BBN, leading to a lower bound of $V_*^{1/4}\gtrsim 10\:\text{MeV}$.

In contrast to slow-roll inflation, chain inflation is not expected to produce any observable tensor modes by the quantum fluctuations of the gravitational field during inflation. This is because the tensor power spectrum is proportional to the overall scale of inflation, $\Delta_t^2\sim V_*/M_P^4$; the bounds of the previous paragraph render the expected value of the tensor-to-scalar ratio $r$ to be too small to be observable. 
In fact, any near-future detection of a non-vanishing $r$ would disfavor chain inflation as a theory of the early Universe.\footnote{In principle, the gravitational waves produced by the first-order phase transitions of chain inflation could also induce a tensor signal in the CMB. However, in the band of extremely-low frequency (aHz$-$fHz) gravitational waves covered by CMB experiments, the signal from the phase transitions is too suppressed to be observable.}

However, we will show that chain inflation, nevertheless, generates strong and potentially testable gravitational radiation. The gravitational waves in chain inflation are not sourced by quantum fluctuations of the gravitational field, but rather they emerge from the first-order phase transitions~\cite{Witten:1984rs,Hogan:1986qda} associated with the quantum tunneling of the inflaton from minimum to minimum along its potential (see also~\cite{Lopez:2013mqa} for the first discussion of looking for gravitational waves from inflation bubble collisions in LIGO --- in the case of a single phase transition). During the phase transitions bubbles of the new vacuum form and expand in the sea of old vacuum~\cite{Coleman:1977py}. Gravitational radiation is produced by the collision of these bubbles~\cite{Kosowsky:1992rz,Kosowsky:1992vn} and by their expansion in the radiation plasma (generated in previous phase transitions) which induces sound waves~\cite{Hindmarsh:2013xza,Hindmarsh:2015qta,Hindmarsh:2017gnf} and magnetohydrodynamic turbulence~\cite{Kosowsky:2001xp,Dolgov:2002ra,Caprini:2009yp}. 

In this work we will derive the full gravitational wave spectrum of chain inflation. For this purpose we will consider an effective parameterization of chain inflation capturing a wide class of models (analogous to the slow-roll expansion in rolling models of inflation). We will show that chain inflation gives rise to spectacular and unique gravitational wave signatures. These include a broad peak in the spectrum from the collection of individual phase transitions. Furthermore, the ``graceful exit'' from chain inflation -- the terminal tunneling of the Universe into its present vacuum -- can induce a second even more intense peak (the latter was originally pointed out in~\cite{Freese:2022qrl}). 

We will investigate in detail the experimental prospects to detect the gravitational wave signal from chain inflation. For this purpose we will calculate the expected signal-to-noise ratio at the most sensitive present and future gravitational wave observatories. We find that pulsar timing array (PTA) experiments are very sensitive to low-scale chain inflation, while space and ground interferometers exhibit a great discovery potential for higher-scale chain inflation. Intriguingly, chain inflation could even be the source~\cite{Freese:2022qrl} of the stochastic gravitational wave background recently observed by the North American Nanohertz Observatory for Gravitational Waves (NANOGrav)~\cite{NANOGrav:2020bcs,NANOGrav:2023gor} - which was confirmed by the European (EPTA)~\cite{Chen:2021rqp,EPTA:2023fyk}, the Parkes (PPTA)~\cite{Goncharov:2021oub,Reardon:2023gzh} and the Chinese Pulsar Timing Array (CPTA)~\cite{Xu:2023wog}.

\section{Chain Inflation}

\subsection{Chain Inflation solves the Empty-Universe problem}\label{sec:emptyuniverse}

In Guth's original inflation model \cite{Guth:1980zm} -- nowadays referred to as old inflation -- the Universe is initially trapped in a false vacuum. Quantum tunneling leads to the formation of true-vacuum bubbles within the sea of false vacuum. In order for the Universe to exit inflation and enter the radiation-dominated epoch, the true vacuum bubbles need to percolate and release their
energy into a hot particle plasma. However, the tunneling rate needs to be sufficiently small for inflation to last long enough to solve, for instance, the horizon problem. As a consequence, in old inflation, the bubbles would form too distantly in space-time to ever collide and the phase transition would never complete~\cite{Guth:1982pn}. Since the insides of the bubbles would forever stay cold and empty in such a universe this issue is referred to as the empty-universe problem of old inflation.

Slow-roll inflation \cite{Linde:1981mu, Albrecht:1982wi} offers a radical solution to the empty-universe problem by entirely dropping the idea of a first-order phase transition and by linking reheating process to the decay of the inflaton (rather than to bubble collisions). There are, on the other hand two different proposals that allow inflation to proceed via first-order phase transitions as in old inflation, yet without ending up with an ``empty universe''. First, the idea of double field inflation~\cite{Adams:1990ds,Linde:1990gz} couples two scalar fields, one rolling and one tunneling field. As the rolling field moves down its potential, it changes the tunneling rate for the other field from initially very slow (allowing sufficient inflation) to suddenly very fast, allowing reheating\footnote{Another way to think about this is in terms of a multidimensional potential, where the field preferentially rolls along one field direction but eventually a rapid tunneling process opens up in another direction.}.

As an alternative, 
chain inflation~\cite{Freese:2004vs,Freese:2005kt,Ashoorioon:2008pj,Winkler:2020ape,Freese:2021noj} remedies the tunneling picture and replaces the single phase transition of old inflation by a series of consecutive phase transitions. The inflaton only resides for a fraction of an e-fold in an individual minimum before it tunnels to the next one along the chain. The fast tunneling ensures that bubbles formed in each phase transition quickly percolate and release their energy into particles. And yet, in the presence of many vacua, inflation lasts long enough to resolve the cosmological conundrums.

In order to make this more explicit let us briefly review the argument by Guth and Weinberg \cite{Guth:1982pn}: in an inflationary universe (with approximately constant Hubble parameter $H$) the volume of an individual vacuum bubble nucleated at the time $t_0$ can be approximated as,
\begin{equation}
V_b \sim \frac{4\pi}{3H^3}e^{3H(t-t_0)}\,,
\end{equation}
where we employed that bubbles approximately expand at the speed of light. Since bubbles are continuously nucleated at the rate per four-volume $\Gamma$, the volume filled by all bubbles compared to the volume of the Universe is
\begin{equation}
\frac{\sum V_b}{V_{\text{tot}}}\simeq \frac{4\pi}{9}\frac{\Gamma}{H^4}\,.
\end{equation}
If this number exceeds unity, the bubbles necessarily percolate and successful reheating is achieved. Hence, the empty-universe problem is evaded if the following condition is met,
\begin{equation}\label{eq:percolation}
\frac{\Gamma}{H^4}>\frac{9}{4\pi}\,.
\end{equation}
This condition implies that the Universe cannot inflate for more than an e-fold in an individual false vacuum.\footnote{This statement holds for a time-independent tunneling rate. It is possible to resolve the empty-universe problem even in a model with only one phase transition through a time-dependent tunneling rate. The latter is realized in double field inflation~\cite{Adams:1990ds,Linde:1990gz}.} Old inflation with a single false vacuum is doomed to fail the percolation condition. However, chain inflation features many successive vacua, along which the inflaton tunnels. Assuming $N_{\text{tot}}\gtrsim 10^2$ vacuum transitions, the Universe only needs to inflate a fraction of an e-fold in each false vacuum, and, yet the total period of inflation can easily exceed the observationally required $20-60$ e-folds. Thus, chain inflation resolves the empty-universe problem.

\subsection{CMB Observables and Implications for the Inflaton Potential}\label{sec:cmb}

In contrast to slow-roll inflation, the origin of density perturbations in chain inflation lies in the stochastic nature of tunneling which induces fluctuations in the number of vacuum transitions and, hence, in the inflaton field value~\cite{Feldstein:2006hm}. We derived the perturbation spectrum in~\cite{Winkler:2020ape}, where we simulated tunneling patterns in the past light cone for a large number of independent realizations. It was found that the scalar power spectrum of chain inflation is determined by the tunneling rate per four-volume $\Gamma$ and the Hubble rate $H$ in the following way,
\begin{equation}\label{eq:powerspectrum}
\Delta_{\mathcal{R}}^2\simeq 0.06 \left(\frac{\Gamma^{1/4}}{H}\right)^{-5/3}\,.
\end{equation}
The scalar power spectrum amplitude is defined as $A_s = \left.\Delta_{\mathcal{R}}^2 \right|_{k=k_*}\,,$
where $k_*=0.05\:\text{Mpc}^{-1}$ denotes the Pivot scale, at which CMB observables are defined. Planck data impose $A_s=(2.10\pm 0.03)\times 10^{-9}$~\cite{Planck:2018jri} which fixes
\begin{equation}\label{eq:GammaH}
\frac{\Gamma^{1/4}_*}{H_*}= 2.98\times 10^4\,.
\end{equation}
Here and in the following, the star indicates that a quantity is evaluated at the moment when the Pivot scale crosses the horizon during inflation.
The simulations also revealed that the lifetime of the Universe in each of the false vacua can be estimated as,
\begin{equation}\label{eq:Deltat}
\Delta t \simeq \frac{1}{1.4 \,\Gamma^{1/4}}\,.
\end{equation}
From Eq.~\eqref{eq:GammaH} and Eq.~\eqref{eq:Deltat} it follows that -- at the time when typical CMB scales cross the horizon -- there need to be around $4.2\times 10^4$ vacuum transitions per e-fold of inflation. While $\Gamma^{1/4}/H$ can significantly change over the course of inflation, the observed near-scale-invariance of the scalar power spectrum within CMB scales implies that $\Gamma^{1/4}/H$ must stay approximately constant at least for a few e-folds. Hence, we can estimate that at least 
\begin{equation}
N_{\text{tot}}>10^5
\end{equation}
vacuum transitions are required for chain inflation to be a viable theory of the early Universe. The correct CMB normalization thus imposes a much stronger constraint on $N_{\text{tot}}$ compared to the percolation condition discussed in the previous section (which would only require $N_{\text{tot}}\gtrsim 100)$. We emphasize, however, that a sizable number of false vacua does not pose any particular model-building challenge and can easily be realized through an axion with a softly broken discrete shift symmetry (as we will discuss in more detail below).

We note that -- due to the large number of phase transitions -- individual contributions to the scalar power spectrum cannot be resolved. Rather, we can treat $\Delta_{\mathcal{R}}^2$ as a continuous and differentiable function of the comoving scale $k$. Similarly, instead of defining a discrete tunneling rate $\Gamma_i$ in each vacuum, we can effectively take a continuum limit and treat $\Gamma$ as a continuous function of the inflaton field vale $\phi$ (or of time $t$) and define the derivatives~\cite{Freese:2021noj},
\begin{equation}\label{eq:derivative}
\Gamma^\prime = \frac{\Delta\Gamma}{\Delta\phi}\,,\qquad
\dot{\Gamma}= \frac{\Delta\Gamma}{\Delta t}\,,
\end{equation}
where $\Delta\Gamma$ is the difference in tunneling rate between two successive vacuum transitions, $\Delta t$ is the time between the transitions (=the vacuum lifetime) and $\Delta\phi$ the field-distance between the two vacua. The derivatives of the Hubble scale $H^\prime$ and $\dot{H}$ can be defined analogously.

Let us now turn to the deviation of the scalar power spectrum of chain inflation from scale-invariance.  As a measure of the deviation from scale-invariance, the following definition of the scalar spectral index (evaluated at the Pivot scale) applies\cite{Winkler:2020ape},
\begin{equation}\label{eq:ns}
n_s = \left. 1+\frac{d\log \Delta_{\mathcal{R}}^2}{d\log k}\right|_{k=k_*} \simeq \left. 1 + \frac{5}{12} \left( \frac{4\dot{H}}{H^2}-\frac{\dot{\Gamma}}{H\Gamma}\right)\right|_{t=t_*}\,,
\end{equation}
with the time-derivatives as defined in Eq.~\eqref{eq:derivative} and $t_*$ denoting the time when the Pivot scale $k_*$ crosses the horizon. Notice that we approximated 
$dk = \dot{k} \, dt \simeq \dot{a}H\, dt = aH^2\, dt$ in the second step, where $a$ is the scale factor. 

In slow-roll inflation the scalar power spectrum typically follows -- to very good approximation -- a power-law form because deviations only enter at higher order in the (small) slow-roll parameters. In contrast, in chain inflation, deviations of $\Delta_{\mathcal{R}}^2$ from the power-law form can occur through the higher time-derivatives $\ddot{\Gamma}$, $\dddot{\Gamma}$ etc.\ which are not necessarily suppressed. As a consequence, the running and running-of-running of the scalar spectral index can be more pronounced in chain inflation. We can express,
\begin{align}\label{eq:runningns}
\frac{d\log n_s}{d\log k} &= \left. \frac{d^2\log \Delta_{\mathcal{R}}^2}{d\log k^2}\right|_{k=k_*}=\left.\frac{5}{12}\left(
-\frac{8\dot{H}^2}{H^4} + \frac{\dot{H}\dot{\Gamma}}{H^3\Gamma} +\frac{\dot{\Gamma}^2}{H^2\Gamma^2}+\frac{4\ddot{H}}{H^3}-\frac{\ddot{\Gamma}}{H^2\Gamma}
\right)\right|_{t=t_*}\,,\\
\frac{d^2\log n_s}{d\log k^2}  &= \left. \frac{d^3\log \Delta_{\mathcal{R}}^2}{d\log k^3}\right|_{k=k_*}=
\frac{5}{12}\left(
\frac{32\dot{H}^3}{H^6}-\frac{3\dot{H}^2\dot{\Gamma}}{H^5\Gamma}-\frac{3\dot{H}\dot{\Gamma}^2}{H^4\Gamma^2}
-\frac{2\dot{\Gamma}^3}{H^3\Gamma^3} 
\right. \nonumber\\
&\hspace{4.2cm} \left. \left. + \frac{\ddot{H}\dot{\Gamma}+3\ddot{\Gamma}\dot{H}}{H^4\Gamma}-\frac{28\ddot{H}\dot{H}}{H^5}+\frac{3\ddot{\Gamma}\dot{\Gamma}}{H^3\Gamma^2}
+\frac{4\dddot{H}}{H^4}-\frac{\dddot{\Gamma}}{H^3\Gamma}
\right)\right|_{t=t_*}\,,
\end{align}
where again all quantities are evaluated at horizon crossing of the Pivot scale.
We note that the above expressions for the running and running-of-running of the spectral index in chain inflation have not previously appeared in the literature.

Since CMB data strongly constrain the amount of running and running-of-running of the spectral index, $\Gamma/H^4$ can maximally change by an $\mathcal{O}(1)$-factor within the entire range of scales accessible in the CMB $k= (10^{-4}\dots 0.5)\:\text{Mpc}^{-1}$. This is remarkable since this range of scales corresponds to $\mathcal{O}(10)$ e-folds of inflation in which $\mathcal{O}(10^5)$ phase transitions need to take place in order to obtain the correct CMB normalization. Hence, we expect changes of only $\Delta \Gamma/\Gamma,\Delta H /H \lesssim 10^{-5}$ between neighboring vacua. A nearly constant tunneling rate implies that the inflaton potential seen from two successive minima needs to look almost identical. In other words, the near scale-invariance of the scalar power spectrum translates to a quasi-periodic shape of the inflaton potential in chain inflation. 

Therefore, the prime inflaton candidate is an axion. At the perturbative level, an axion $\phi$ possesses an (exact or approximate) shift symmetry $\phi\rightarrow \phi + c$, which is broken by non-perturbative instanton effects. The latter induce a periodic potential -- in the simplest case a cosine potential -- with the axion decay constant $f$ defining its periodicity. In addition, a small explicit symmetry violation can break the degeneracy between minima along the periodic potential.\footnote{Consider, for instance, the case where the axion is the pseudo Nambu-Goldstone boson associated with the spontaneous breaking of a global U(1)-symmetry. Because global symmetries in nature are likely not exact, a quasi-periodic (rather than a periodic) axion potential is expected to emerge~\cite{Holman:1992us,Kamionkowski:1992mf,Barr:1992qq,Ghigna:1992iv}.} Locally, we may thus approximate $V(\phi)$ by a tilted cosine,
\begin{equation}\label{eq:tiltedcosine}
V(\phi) \simeq -\mu^3\phi + \Lambda^4\cos\left(\frac{\phi}{f}\right) + C\,,
\end{equation}
where the coefficients $\mu$, $\Lambda$, $f$, $C$ are virtually constant if we zoom in on a few neighboring minima in the potential. However, over the entire field range traversed during inflation (which spans $\gtrsim 10^5$ local minima) they may somewhat vary (the variation is, however, limited by the mentioned constraints on the running and running-of-running of the spectral index).\footnote{Such slow variations of the coefficients naturally arise in a full model realization, where the coefficients often depend on the vacuum expectation values of some heavy fields which are integrated out. These vacuum expectation values may change slowly during inflation due to couplings of the inflaton to the heavy fields (see e.g.~\cite{Flauger:2014ana})} 

The inflaton potential in Eq.~\eqref{eq:tiltedcosine} has been considered for slow-roll inflation in the regime, where the potential is monotonic~\cite{Silverstein:2008sg,McAllister:2008hb} (see also~\cite{Kappl:2015esy,Winkler:2019hkh}). In contrast, for $\Lambda^4 > \mu^3 f$, it features a series of metastable minima and is suitable for chain inflation. In~\cite{Freese:2021noj} we showed that -- as long as chain inflation is locally described by Eq.~\eqref{eq:tiltedcosine} -- an upper limit on the scale of inflation at horizon crossing of CMB scales arises,
\begin{equation}\label{eq:maxscale}
V_*^{1/4}< 10^{12}\:\text{GeV}\,.
\end{equation}
We obtained this bound by imposing a combination of constraints: requiring the correct normalization and spectral index of the scalar power spectrum; perturbative unitarity; requiring consistent tunneling solutions and avoiding a tunneling catastrophe.
Deviations of the inflaton potential from Eq.~\eqref{eq:tiltedcosine} -- for instance invoking a more general periodic function instead of the cosine -- may somewhat alter the bound on the scale of inflation. However, even in this case, the derivation in~\cite{Freese:2021noj} applies with minor modifications. Therefore, we do not expect viable chain inflation models with an inflation scale considerably above $V_*^{1/4}= 10^{12}\:\text{GeV}$.

Let us finally remark that, as a consequence, we expect the tensor-to-scalar ratio $r$ of chain inflation to be highly suppressed. While a complete derivation of the tensor power spectrum of chain inflation is still missing in the literature, we can employ the fact that fluctuations in the metric should scale with the overall energy density. Therefore, the amplitude of tensor modes due to quantum fluctuations will be controlled by the scale of inflation (as in slow-roll inflation). The constraint on $V_*$ in Eq.~\eqref{eq:maxscale}, hence, suggests the absence of observable tensor modes from chain inflation. We note that chain inflation, in principle, also produces tensor modes by the bubble collisions during the first-order phase transitions. The so-produced gravitational waves do not fall into the sensitivity window of CMB experiments (i.e.\ they do not contribute to $r$ which is defined at the Pivot scale). Yet, we will show in Sec.~\ref{sec:SearchingGW} that the bubble-induced gravitational waves give rise to spectacular signatures at interferometer and PTA experiments.

\subsection{On the Graceful Exit from Chain Inflation}\label{sec:exit}

In the previous section we found that viable chain inflation features a large number ($N_{\text{tot}}\gtrsim 10^5$) of vacuum transitions, each of which proceeds in a small fraction of a Hubble time. However, eventually, a ``graceful exit'' from chain inflation is required: the vacuum tunneling needs to stop and the Universe must enter the radiation-dominated epoch. While the generation of a radiation bath is automatically achieved by the bubble collisions during the phase transitions -- as long as the percolation condition is satisfied -- the successful termination of chain inflation is less trivial. In particular, it must be ensured that no vacuum transitions occur after primordial nucleosynthesis (BBN) in order not to spoil the light element abundances. 

\begin{figure}[htp]
\begin{center}
  \vspace{-5mm}
  \includegraphics[width=11.5cm]{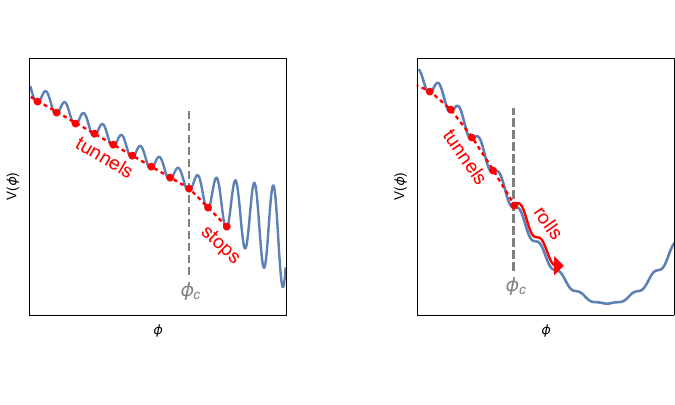}
\end{center}
\vspace{-1.2cm}
\caption{Illustration of two scenarios for a graceful exit from chain inflation. In both cases the inflaton tunnels from minimum to minimum along a quasi-periodic potential (from left to right in the figures). But once it reaches a critical field value $\phi_c$ chain inflation ends and the Universe enters the radiation-dominated epoch. In scenario~1 (left panel) this is achieved by an increase in the potential barriers occurring at $\phi>\phi_c$. The lifetime of the vacua after $\phi_c$ successively increases and, once it exceeds the age of the Universe, the inflaton remains trapped. In contrast, in scenario~2 (right panel) the potential barriers become shallower and shallower until -- at the critical field value $\phi_c$ -- the potential becomes monotonic. Hence, at $\phi=\phi_c$ the inflaton switches from tunneling to rolling down the potential. Similar as in slow-roll inflation the field then oscillates around the global minimum  of its potential before it decays away into radiation.} 
\label{fig:graceful}
\end{figure}

Two key scenarios of a graceful exit from chain inflation are illustrated in Fig.~\ref{fig:graceful}. In the scenario~1 (left panel) the inflaton tunnels quickly from minimum to minimum along its potential. But once it reaches a critical field value $\phi_c$, the barriers in the potential increase and the tunneling rate -- which is exponentially sensitive to the barrier height -- becomes suppressed. Hence, inflation ends around $\phi_c$ in a metastable vacuum which hosts the Universe until today (but which will eventually decay in the far future). An important aspect is that the crossover from fast tunneling during inflation to metastability after inflation needs to occur rapidly -- typically within a few phase transitions -- in order to avoid that any transitions occur after BBN. We also note that the tunneling rate is exponentially sensitive to potential parameters, so that small parameter changes lead to very different tunneling rates.

A simple model realization of scenario~1 -- inspired by the relaxion mechanism~\cite{Graham:2015cka} -- has been given in~\cite{Freese:2021rjq,Freese:2022qrl}. It employs a second auxiliary scalar field $\chi$ coupled to the inflaton which plays no role during inflation, but triggers a rapid increase of the potential barriers in $\phi$-direction once the inflaton passes a critical field value $\phi_c$. The inflaton itself resides in a tilted cosine-potential,
\begin{equation}\label{eq:chainmodel}
 V = -\mu^3\phi + \Lambda^4\cos\left(\frac{\phi}{f}\right) + C  + V_{\text{coupling}}\,,
\end{equation}
where $\mu$, $\Lambda$ and $f$ and $C$ may either be constants or slowly varying functions of $\phi$ (see Sec.~\ref{sec:cmb}). For $\Lambda > \mu^3\,f$ the potential exhibits a series of unstable minima suitable for chain inflation. The coupling terms read
\begin{equation}\label{eq:relax}
 V_{\text{coupling}} = (M_1^2 - M_2 \phi)\chi^2 + {\Lambda^\prime}^2 \chi^2\cos\frac{\phi}{f} + \lambda\chi^4 \,,
\end{equation}
where $M_{1,2}$ and $\Lambda^\prime$ denote mass parameters, while $\lambda$ stands for the self-coupling of $\chi$. 

Chain inflation starts at $\phi\ll 0$, while $\chi$ is strongly stabilized at the origin in field-space such that we can effectively set $V_{\text{coupling}}=0$, and Eq.~\eqref{eq:chainmodel} reduces to Eq.~\eqref{eq:tiltedcosine}. The inflaton tunnels from wiggle to wiggle along the tilted cosine potential. However, once $\phi$ reaches the critical field value $\phi_c\simeq M_1^2/M_2$, the $\chi^2$-term in Eq.~\eqref{eq:relax} turns negative and $\chi$ gets displaced. Thereby, it quickly raises the potential barriers in $\phi$-direction (via the $\Lambda^\prime$-term in Eq.~\eqref{eq:relax}) such that tunneling slows down and then stops (within the lifetime of the Universe) after a few more transitions. For instance, in the parameter example presented in~\cite{Freese:2022qrl}, the inflaton undergoes $\sim 10^6$ tunnelings with $\Delta t\simeq 0.6\:\text{ns}$ during inflation. But once it passes the critical field value where the potential barriers increase, only two more transitions with $\Delta t\simeq 0.7\:\text{ns}$ and $\Delta t\simeq 0.01\:\text{s}$ occur, while the next vacuum is already stable within the age of the Universe. Such a sharp increase in lifetime within a few transitions may seem surprising. But due to the exponential sensitivity of the lifetime on the potential parameters it corresponds to an increase of the height of the potential barriers by only a factor $\sim 2$. 

In the second scenario (right panel of Fig.~\ref{fig:graceful}) chain inflation also proceeds by tunneling along a (quasi)periodic potential. However, in this case, the potential barriers become shallower and shallower until, at a critical field value $\phi_c$, the potential becomes monotonic and the inflaton quickly starts rolling down the potential.\footnote{Shortly before the inflaton reaches the critical field value -- when the potential barriers become small compared to the energy density difference between minima -- there may also occur a ``tunneling catastrophe'' in which the bubble wall collisions trigger a chain reaction of new vacuum transitions such that the inflaton effectively directly tunnels to the bottom of the potential rather than into the neighboring metastable minimum~\cite{Cline:2011fi}. The occurrence of such a tunneling catastrophe, however, depends on details of the underlying model, in particular the efficiency of vacuum energy transfer into radiation~\cite{Easther:2009ft,Giblin:2010bd}. Since in both cases -- direct tunneling or fast rolling to the bottom of the potential -- chain inflation ends around $\phi\sim \phi_c$, they produce similar CMB observables. Hence, we do not need to consider the tunneling catastrophe as a separate scenario.}  
Since the potential is typically too steep to feature any slow-roll regime\footnote{\label{foot:noslowroll}We remind the reader that chain inflation does not require an extremely flat potential (in contrast to slow-roll inflation). It is, hence, unlikely that chain inflation features a slow-roll regime in its potential just by chance.} inflation immediately ends at $\phi=\phi_c$. The inflaton quickly rolls towards and then oscillates around the global minimum in its potential, before it decays away into radiation. Scenario~2 for the graceful exit is, for instance, realized in natural chain inflation~\cite{Freese:2021noj} which features the following inflaton potential,
\begin{equation}\label{eq:potentialmodel}
V= V_0 \left[1-\cos\left(\frac{\phi}{F}\right)+\delta\sin\left(\frac{\phi}{2F}\right)\sin\left(\frac{\phi}{f}\right)\right]\,.
\end{equation}
Such a potential was shown to naturally arise for axions in supergravity~\cite{Freese:2021noj}. Given $F\gg f$, $\delta\ll 1$, it supports a regime of chain inflation which is terminated at the critical field value $\phi_c= 2 F \arccos(F\delta/2f)$, where the inflaton starts rolling as described above.

Let us finally note that the radiation plasma, which is continuously populated during chain inflation by the phase transitions, can also participate in the graceful exit. This is because a decrease (increase) in the temperature of the radiation may slow-down (speed-up) the phase transitions and, therefore, impact the mechanism by which inflation stops. However, the thermal effect on the tunneling rate depends on details of the underlying inflation model, for instance on
the inflaton-radiation coupling.

\section{Evolution of Energy Densities}\label{sec:evolution}

Chain inflation is realized in a wide class of models with quasi-periodic inflaton potentials. The inflaton tunnels from minimum to minimum along the potential, where the transition rate between two minima is given by~\cite{Coleman:1977py,Callan:1977pt,Linde:1980tt,Linde:1981zj},
\begin{equation}\label{eq:transitionrate}
 \Gamma \simeq \text{max}\left[m^4 \left(\frac{S_{E}}{2\pi}\right)^2 e^{-S_{E}},\:
 T^4 \left(\frac{S_{E}}{2\pi\,T}\right)^{3/2} e^{-S_{E}/T} \right]\,.
\end{equation}
Here $S_{E}$ stands for the Euclidean action of the bounce solution extrapolating between the two vacua, while $m$ is the mass of the tunneling scalar field (evaluated in the vacuum populated before the transition). The first term in~Eq.~\eqref{eq:transitionrate} corresponds to the quantum tunneling rate at zero temperature, while the second term is the thermally-induced rate. In the absence of a radiation plasma -- this can be the case for instance at the beginning of chain inflation -- $\Gamma$ is always given by the quantum tunneling rate. If a plasma with temperature $T$ is present, it depends on the specific model realization which of the two rates dominates. In this case $\Gamma$ is determined by the faster of the two rates. 

The observed normalization and near scale-invariance of the scalar power spectrum imply that chain inflation consists of many quick transitions among which $\Gamma$ changes very slowly. Therefore, as we noted previously, instead of considering $\Gamma$ to be a discrete quantity which changes from vacuum to vacuum we can effectively take it to be a continuous function of the inflaton field value (or of the time). Correspondingly, we can define the field- and time derivatives of $\Gamma$ as in Eq.~\eqref{eq:derivative} (and analogously of the Hubble scale $H$). 

In order to calculate the evolution of the vacuum and radiation energy densities $\rho_{\text{vac}}$ and $\rho_{\text{rad}}$ during inflation, we can thus take the phase transitions to continuously convert vacuum energy into radiation energy, the latter of which is also subject to redshifting. We arrive at the following set of differential equations,
\begin{equation}\label{eq:rhodgl}
\rho^\prime_{\text{vac}}(\phi)= -\frac{\Delta V}{\Delta\phi}\,,\quad\qquad
\rho^\prime_{\text{rad}}(\phi)= 
\frac{\Delta V}{\Delta\phi} - \frac{4\, H  \rho_{\text{rad}}\,\Delta t}{\Delta\phi}
\simeq
\frac{\Delta V}{\Delta\phi} - \frac{4\, H  \rho_{\text{rad}}}{1.4\,\Gamma^{1/4}\Delta\phi}
\,,
\end{equation}
where we used Eq.~\eqref{eq:Deltat} to express $\Delta t$ in terms of $\Gamma$. In the above set of equations we treat all quantities as continuous functions of $\phi$ which is justified by the many transitions which occur per Hubble time.\footnote{Alternative to Eq.~\eqref{eq:rhodgl} one can employ a discrete approach, where one treats every phase transition as a discrete event. We have verified, however, that for all viable chain inflation models both approaches yield equivalent density evolutions (with the discrete approach requiring significantly more computational resources).} The Hubble rate is determined by the energy densities,
\begin{equation}\label{eq:Hubble}
H=\sqrt{\rho_{\text{vac}}(\phi)+\rho_{\text{rad}}(\phi)}/(\sqrt{3}\,M_P)\,.
\end{equation}

For an explicit choice of the inflaton potential the density evolution can be derived by calculating $\Gamma$ in each vacuum along the chain via Eq.~\eqref{eq:transitionrate}. At each tunneling step this requires the determination of the Euclidean action of the bounce -- either by solving the differential equation of the bounce, or by employing an analytic approximation. For instance, the thin-wall approximation of vacuum tunneling predicts~\cite{Coleman:1977py},
\begin{equation}\label{eq:thinwall}
S_E = \frac{27\pi^2\mathcal{S}^4}{2(\Delta V)^3}\qquad \text{with}\qquad \mathcal{S}=\int\limits_{\phi_i}^{\phi_{i+1}}d\phi \sqrt{2 V(\phi)-2 V(\phi_{i+1})}\,\sim \sqrt{V_b} \:\Delta\phi \,,
\end{equation}
where $\phi_i$ and $\phi_{i+1}$ denote the field values and $\Delta\phi=\phi_{i+1}-\phi_i$ the field-distance of the two neighboring minima in the potential between which the inflaton tunnels. The height of the potential barrier is denoted by $V_b$. Validity of the thin-wall approximation requires $\Delta V\ll V_b $ which is hardly ever achieved in realistic tunneling phenomena.\footnote{In the regime $\Delta V\ll V_b $ the lifetime of the false vacuum is typically longer than the age of the Universe.} However, correction factors which generalize Eq.~\eqref{eq:thinwall} outside the thin-wall regime have been obtained for several standard potentials (including tilted-cosine potentials~\cite{Winkler:2020ape} and quartic potentials~\cite{Adams:1993zs}). Once $\Gamma$ at the field-values of the subsequent minima has been determined, we can interpolate it to obtain a continuous function $\Gamma(\phi)$. The latter is then plugged into Eq.~\eqref{eq:rhodgl} in order to derive the evolution $\rho_{\text{vac}}(\phi)$, $\rho_{\text{rad}}(\phi)$ which also fixes $H(\phi)$ through Eq.~\eqref{eq:Hubble}. The evolution of $\Gamma(\phi)$ and $H(\phi)$ then fix the scalar power spectrum $\Delta_{\mathcal{R}}^2(\phi)$ as a function of the inflaton field-value through Eq.~\eqref{eq:powerspectrum}. In order to express the power spectrum in terms of the comoving scale $k$ -- as required for the comparison with observation -- we can employ the horizon crossing condition,
\begin{equation}\label{eq:matching}
k(\phi)=a(\phi)H(\phi)=a_*e^{\mathcal{N}(\phi)}H(\phi)\qquad\text{with}\qquad \mathcal{N}(\phi) = \int\limits_{\phi_*}^{\phi}d\phi^\prime H(\phi^\prime)(\Delta t/\Delta\phi)\,,
\end{equation}
where $\mathcal{N}(\phi)$ denotes the number of e-folds of expansion, while the inflaton travels from $\phi_*$ to some field-value $\phi$. The scale factor at the Pivot scale $a_*$ is determined by $a_*H_*=k_*$. For consistency, we must also ensure that the so-determined value of $a_*$ is consistent with $a_0=1$, where $a_0$ is the scale factor in today's Universe. For evaluating this condition let us turn to the end of inflation.

The precise definition of the end of chain inflation is somewhat ambiguous. This is because -- towards the end of chain inflation -- there occurs a (relatively short) time period at which the Universe is already radiation-dominated, but the phase transitions still continue for some time until all vacuum energy has been released. It is a matter of convention if one considers this period as a part of inflation. As long as one consistently employs the same definition of the end of chain inflation, observables are not affected by this choice.

In Sec.~\ref{sec:exit} we described the two scenarios of a graceful graceful exit from chain inflation. In both scenarios, there exists a critical inflaton field value $\phi_c$ at which the dynamics of inflation changes -- either by a sudden increase in the potential barriers which quickly stops the tunneling (scenario~1) or by a crossover from tunneling to rolling (scenario~2). In the following we will define the time $t_c$ when the inflaton reaches the value $\phi_c$ as the end of inflation. This definition turns out to be convenient because until $t_c$, many phase transitions occur per Hubble time. In this regime we can consistently work in the ``continuum limit'' where we treat $\Gamma$ as continuous in $\phi$ and derive the density evolution through~Eq.~\eqref{eq:rhodgl}. There may occur up to a handful of ever slower phase transitions (in scenario~1) after $t_c$ which -- as we will see -- have rather different properties compared to the quick transition during chain inflation. Therefore, we do not consider them as part of inflation.

We note that radiation-domination already sets in slightly before $t_c$. This is because it takes about an e-fold of inflation to redshift away significant amounts of radiation. During inflation, there is always a radiation background with an energy density approximately corresponding to the vacuum energy released during one e-fold of inflation. Hence, the Universe becomes radiation-dominated $\mathcal{O}(1)$ e-fold before $t_c$. The radiation produced during the last e-fold of inflation makes up most of the radiation we observe in the Universe today (unless post-inflationary entropy production takes place and additionaly contributes to the radiation bath). Since there generically occur at least $\mathcal{O}(10^3)$ phase transitions in the last e-fold, we can estimate $\rho_{\text{rad},c} \gtrsim 10^3\,\Delta V \gg \rho_{\text{vac},c}$ (where we employed that the radiation produced during the last e-fold has not strongly redshifted by the end of inflation).
Here and in the following the subscript $c$ indicates that a quantity is evaluated at $t_c$. The temperature $T_c$ of the radiation bath at $t_c$ can be called the reheating temperature of the Universe. Neglecting the subdominant vacuum contribution we determine $T_c$ by the relation
\begin{equation}
\rho_{\text{rad},c} \simeq \frac{\pi^2}{30}\,g_{\text{eff}}(T_c) \,T_c^4\,,
\end{equation}
where $g_{\text{eff}}$ stands for the number of relativistic degrees of freedom in the radiation plasma. Furthermore, we can define the entropy density at the end of chain inflation,
\begin{equation}
s_c \simeq\frac{2\pi^2}{45}\,g_{\text{eff}}(T_c) \,T_c^3\,.
\end{equation}

If the Universe subsequently runs through a standard evolution with radiation-domination followed by matter-domination, the entropy of the Universe is approximately conserved. However, depending on how quickly the remaining vacuum energy $\rho_{\text{vac},c}$ is released, there may also occur a second short period of vacuum domination long after $t_c$. In this case an additional entropy increase occurs in the final phase transition(s) when $\rho_{\text{vac},c}$ is transformed into radiation.\footnote{Of course any phase transition after $t_c$ increases the entropy of the Universe. However, if the vacuum energy is (strongly) subdominant, this entropy release is negligible.} In order to cover both cases we define,
\begin{equation}\label{eq:entropyfactor}
s_0 = s_c\, a_c^3 \, \Delta_\xi\,,
\end{equation}
where the entropy factor $\Delta_\xi=1$ if entropy is conserved, while $\Delta_\xi>1$ in the presence of entropy production after $t_c$ (i.e.\ in the case of a second period of vacuum-domination). The subscript $0$ refers to today's universe.

With these definitions we can write the matching condition for the scale factor in the following way,
\begin{equation}\label{eq:astar}
 a_*  = a_c\, e^{-\mathcal{N}(\phi_c)} = \left(\frac{g_{\text{eff}}(T_0)}{g_{\text{eff}}(T_c)}\right)^{1/3}\left(\frac{T_0}{T_c}\right)\,\Delta_\xi^{-1/3}\,e^{-\mathcal{N}(\phi_c)}
\end{equation}
where, in the last step, we employed Eq.~\eqref{eq:entropyfactor} and used that $s(T)\propto g_{\text{eff}}(T)\,T^3$.
In a consistent model of chain inflation, the parameters must be adjusted in such a way that $a_*$ fulfills the above matching condition at the Pivot scale $k_*=a_* H_*$. Depending on the scale of inflation this is typically achieved for $\mathcal{N}(\phi_c)=20-50$.

\section{An Effective Theory of Chain Inflation}\label{sec:parameterization}

While the procedure of the previous section allows for the calculation of the CMB observables in any concrete model, the focus of this work is not on a specific chain inflation realization. Instead, we want to investigate the gravitational wave signals in a wide class of chain inflation models. Therefore, rather than specifying a concrete inflaton potential, we will introduce an effective parameterization which covers a large space of possible models. 

\subsection{Parameterizing Chain Inflation}

Specifically, we will employ the fact that the near scale-invariance of the scalar power spectrum requires the tunneling rate to only change very slowly along the different minima in field space (see Sec.~\ref{sec:cmb}). Variations of the action $S_E$ are even more constrained due to the exponential sensitivity $\Gamma\propto e^{-S_E}$. Therefore, we perform an expansion of the Euclidean action around the field-value $\phi_*$, at which CMB scales (more precisely the Pivot scale) cross the horizon,
\begin{equation}\label{eq:Sexpansion}
S_E(\phi) = S_{E,*}+ S_1 \left(\frac{\phi-\phi_*}{\Delta\phi_{\text{tot}}}\right)+ S_2 \left(\frac{\phi-\phi_*}{\Delta\phi_{\text{tot}}}\right)^2+ S_3 \left(\frac{\phi-\phi_*}{\Delta\phi_{\text{tot}}}\right)^3+\dots\,,
\end{equation}
where $\Delta\phi_{\text{tot}}$ is the field range traversed between horizon crossing of the CMB scales and the end of inflation (more precisely between horizon crossing of the Pivot scale and the end of inflation). The above expansion will turn out to be very useful because we can use CMB data to constrain the parameters $S_i$. This will later allow us to estimate the strength of the gravitational wave signal from chain inflation. 
Notice that we picked the normalization of the coefficients $S_i$ such that $\sum_i S_i = \Delta S$ is the difference in the bounce action between horizon crossing of the CMB scales and the end of inflation.

Notice that the expansion is meant to describe the case where (zero-temperature) vacuum transitions dominate the tunneling rate. The slow variation of $S_E$ along vacua can be caused by gradual changes of $\Delta V$, $\Delta \phi$ and $V_b$. In the case of thermal transitions it is more convenient to expand $S_E/T$ instead of $S_E$ since this is the quantity which appears in the exponent of the transition rate. But otherwise the following discussion identically applies to the thermal case. Hence, all the results of this work are valid independent of whether (zero-temperature) vacuum transitions or thermal transitions dominate during chain inflation.

We decided to consider the expansion up to 4-th order, i.e.\ we will include the terms from $S_1$ to $S_4$. Including even more higher-order terms does not lead to qualitatively new chain inflation phenomenology. Furthermore, we emphasize that the expansion in Eq.~\eqref{eq:Sexpansion} is only applicable if the higher order terms are suppressed against the zeroth order term, i.e.\ if $|S_i| < S_{E,*}$. We verified that the CMB constraints on the scale-dependence of the scalar power spectrum automatically enforce this condition. Therefore, the interesting chain inflation model space indeed maps to the regime where the expansion is meaningful (see Sec.~\ref{sec:scaledependence}).

With the above said, the expansion of the Euclidean action of the bounce translates to the following expansion of the vacuum transition rate,
\begin{equation}\label{eq:Gammaparametrization}
\Gamma = \Gamma_* \:e^{ -S_1 \,\tilde{\phi}- S_2 \,\tilde{\phi}^2- S_3\, \tilde{\phi}^3 - S_4 \,\tilde{\phi}^4}\qquad\text{with}\qquad
\tilde{\phi} = \frac{\phi-\phi_*}{\Delta\phi_{\text{tot}}}
\,,
\end{equation}
where we neglected subdominant changes of $\Gamma$ caused by the $S_E$-dependent prefactor in Eq.~\eqref{eq:transitionrate}. Notice that, for convenience, we introduced the rescaled field value $\tilde{\phi}$ which evolves from $\tilde{\phi}=0$ at horizon crossing of the Pivot scale to $\tilde{\phi}=1$ at the end of inflation. The tunneling rate at the end of inflation is given by $\Gamma_c=\Gamma_*\,e^{-S_1-S_2-S_3-S_4}$.

Apart from $\Gamma$, the chain inflation dynamics depend on the evolution of $\Delta V$ and $\Delta \phi$ (cf.\ Eq.~\eqref{eq:rhodgl}). In fact, $S_E$ and $\Delta V$, $\Delta \phi$ are correlated since the shape of the potential determines the Euclidean action. Nevertheless, it turns out to be a good approximation to treat $\Delta V$ and $\Delta \phi$ as constants. This is because the transition rate $\Gamma$ is extremely sensitive to small changes in the potential. 
For instance, if we apply the thin-wall approximation (Eq.~\eqref{eq:thinwall}), we find that a $10\%$ change of $\Delta V$ or $\Delta\phi$ over the course of inflation alters $\Gamma$ by $5-35$ orders of magnitude.\footnote{While most chain inflation models do not fall into the thin-wall regime, the thin-wall approximation, nevertheless, reasonably captures the scaling of $S_E$ with $\Delta V$ and $\Delta\phi$. For arriving at the above estimate we also took into account that $S_{E,*}\simeq 30-120$ in viable chain inflation models.} Therefore, the dynamics of chain inflation and the resulting CMB observables are controlled by the evolution of $\Gamma$, while we can neglect\footnote{More precisely, we neglect the direct impact of the field-dependence of $\Delta V$, $\Delta\phi$. The fact that $\Gamma$ changes over the course of inflation due to changes of $\Delta V$, $\Delta\phi$ is taken into account by the expansion in Eq.~\eqref{eq:Gammaparametrization}.} the (tiny) field-dependence of $\Delta V$ and $\Delta\phi$. With this simplification, the differential equations for the density evolutions (cf. Eq.~\eqref{eq:rhodgl}) can be rewritten as,
\begin{equation}\label{eq:rhodgl2}
\frac{d \rho_{\text{vac}}}{d\tilde{\phi}}= -\Delta V_{\text{tot}}\,,\qquad\quad
\frac{d \rho_{\text{rad}}}{d\tilde{\phi}}\simeq  \Delta V_{\text{tot}} \,\left(1-\frac{4\, H  \, \rho_{\text{rad}}}{1.4\,\Delta V\,\Gamma_*^{1/4} }\,e^{ (S_1 \,\tilde{\phi}+ S_2 \,\tilde{\phi}^2+ S_3\, \tilde{\phi}^3 + S_4 \,\tilde{\phi}^4)/4}\right)
\,,
\end{equation}
where we plugged in our parameterization of the tunneling rate in Eq.~\eqref{eq:Gammaparametrization}. Furthermore, we introduced the total vacuum energy released between horizon crossing of the Pivot scale and the end of inflation $\Delta V_{\text{tot}}$ and used,
\begin{equation}
    N_{\text{tot}} = \Delta\phi_{\text{tot}}/\Delta\phi=\Delta V_{\text{tot}}/\Delta V\,,
\end{equation}
with $N_{\text{tot}}$ denoting the total number of phase transitions 
between horizon crossing of the Pivot scale and the end of inflation. 
Notice that the equation for the vacuum energy has the following simple analytic solution,
\begin{equation}
\rho_{\text{vac}}(\tilde{\phi}) = V_* - \Delta V_{\text{tot}}\;\tilde{\phi}\,,
\end{equation}
while the differential equation for the radiation density must be solved numerically. We note that -- quickly after the start of chain inflation -- the radiation density approaches an attractor solution which features an equilibrium between radiation production (by the phase transition) and redshifting.  Hence, after a few e-folds of inflation, $\rho_{\text{rad}}$ becomes insensitive to its initial value. Assuming that the attractor solution is reached before the observable inflation starts we can estimate the radiation density $\rho_{\text{rad},*}$ at horizon crossing of the Pivot scale by solving the implicit equation,
\begin{equation}\label{eq:rhoradstar}
 \rho_{\text{rad},*} \simeq \frac{1.4\, \Delta V \: \Gamma_*^{1/4}}{4H_*}\qquad\text{with}\qquad H_* = \sqrt{\frac{V_*+\rho_{\text{rad},*}}{3\,M_P^2}}\,,
\end{equation}
where we employed Eq.~\eqref{eq:rhodgl2} and approximated 
$d\rho_{\text{rad}}/d\tilde{\phi}\simeq 0$ at $t_*$. The radiation energy density $\rho_{\text{rad},*}$ is of the order of the vacuum energy released during one e-fold of inflation. This is because it takes $\mathcal{O}(1)$ e-fold to redshift away the radiation from previous phase transitions.  

The expansion in Eq.~\eqref{eq:Sexpansion} can be thought of in a sense as similar to the expansion in terms of slow-roll parameters $\epsilon$ and $\eta$ in slow-roll inflation. In the case of chain inflation, the CMB observables are more related to the tunneling rate. 
Due to the exponential dependence on the Euclidean action, small changes in the potential lead to huge changes in the tunneling rate and therefore in CMB observables.  For the case of a pure tilted cosine, all  $S_i =0$ and $\Gamma = \Gamma_*$. However, any addition to the tilted cosine such as an additional small quadratic term would yield nonzero $S_i$. 

\subsection{Parameterizing the Graceful Exit}

Finally, we need to model the end of inflation and the transition into the radiation-dominated epoch. In Sec.~\ref{sec:exit} we described two scenarios of a graceful exit from chain inflation, in which inflation ends once the inflaton reaches a critical field value $\phi_c$. In scenario~1, this is achieved by a quick increase of the potential barriers -- for instance triggered by an auxiliary field which receives a vacuum expectation value -- which quickly stops the tunneling. In this case, a handful of ever slower transitions may occur at $\phi>\phi_c$ until the inflaton settles in a minimum with a lifetime larger than the age of the Universe. We have verified that, if several (ever slower) phase transitions occur after $\phi_c$, the entropy increase and the expected gravitational wave signal are completely dominated by the last phase transition. Because only the last transition affects cosmological observables, it is sufficient to model this last transition in our parameterization of the graceful exit. In scenario~2 the inflaton starts rolling (or directly tunnels to the bottom of the potential) at $\phi_c$ and inflation ends. By considering either one or zero phase transition(s) after inflation, we, therefore, effectively capture all relevant scenarios of a graceful exit.

We will thus apply the following two boundary conditions for the vacuum energy at the end of inflation:
\begin{enumerate}[(i)]
    \item $\rho_{\text{vac},c} = \Delta V\,$,
    \item $\rho_{\text{vac},c} = 0\,$,
\end{enumerate}

where the condition (i) corresponds to the case of one post-inflationary phase transition (which also captures the case of several post-inflationary transitions as explained above). This condition, hence, applies to scenario~1 of Fig.~\ref{fig:graceful}. Condition (ii) corresponds to the case with no post-inflationary transition as in scenario~2 of Fig.~\ref{fig:graceful}.\footnote{We note that scenario~1 of Fig.~\ref{fig:graceful} can also give rise to the boundary condition (ii). This happens if the potential barriers increase extremely rapidly for $\phi>\phi_c$ such that the tunneling immediately stops when the inflaton reaches the field-value $\phi_c$, i.e.\ not a single further phase transition occurs within the lifetime of the Universe.} Setting the vacuum energy to zero (i.e.\ assuming instantaneous reheating) for this case is a slight oversimplification which does, however, qualitatively not affect our results (see also footnote~\ref{foot:noslowroll}).

In case (i), we also need to specify when the post-inflationary transition occurs. Instead of stating the time $t_{\text{last}}$ (or the Euclidean action of the bounce) associated with this last transition it is more convenient to specify the ratio of vacuum energy to radiation energy right before the transition~\cite{Kamionkowski:1993fg},
\begin{equation}\label{eq:alpha}
    \alpha = \left. \frac{\rho_{\text{vac}}}{\rho_{\text{rad}}}\right|_{t=t_{\text{last}}}\,,
\end{equation}
which is also called the strength of the phase transition. While the Universe is radiation-dominated at the end of inflation, there may occur a later (second) period of vacuum-domination if the last transition occurs relatively late. In this case, values of $\alpha$ larger than one are realized. The second vacuum-domination can, however, only last for a short period (at most a fraction of an e-fold) in order not to violate the percolation condition (see Sec.~\ref{sec:emptyuniverse}). Correspondingly an upper bound $\alpha < 21$ arises~\cite{Freese:2022qrl}. In a concrete model realization, the value of $\alpha$ is determined by the stopping mechanism which terminates inflation (see~\cite{Freese:2022qrl}).

In total, our model-independent parameterization employs seven independent input parameters to describe the dynamics of chain inflation. These are the energy scale of inflation $V_*$, the energy difference between vacua $\Delta V$, the vacuum transition rate at horizon crossing of the CMB scales $\Gamma_*$ and the parameters $S_{1,2,3,4}$ which determine the evolution of the vacuum transition rate during chain inflation. In addition, we need one parameter $\alpha$ to characterize the graceful exit from chain inflation. In the absence of any post-inflationary phase transitions we set $\alpha=0$. If there occurs one post-inflationary transition (which also effectively captures the case of several post-inflationary transitions, see above) $\alpha$ takes a non-zero value and determines the vacuum-to-radiation energy ratio right before the transition (cf.\ Eq.~\eqref{eq:alpha}). The latter fixes the time of the last transition.\footnote{Of course we could equally well use the time or the Euclidean action of the post-inflationary transition as an input parameter and take $\alpha$ to be a derived parameter.} 

\begin{figure}[t]
    \centering
    \includegraphics[width=0.49\textwidth]{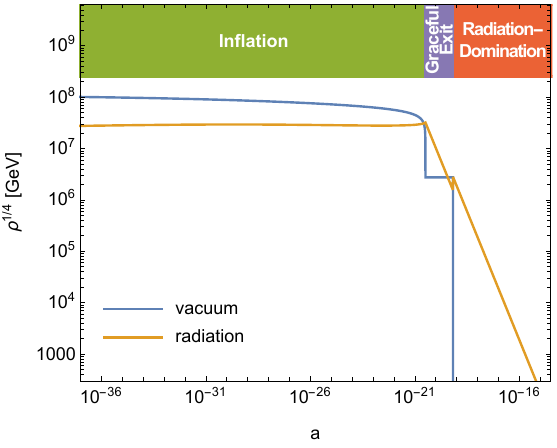}
    \includegraphics[width=0.49\textwidth]{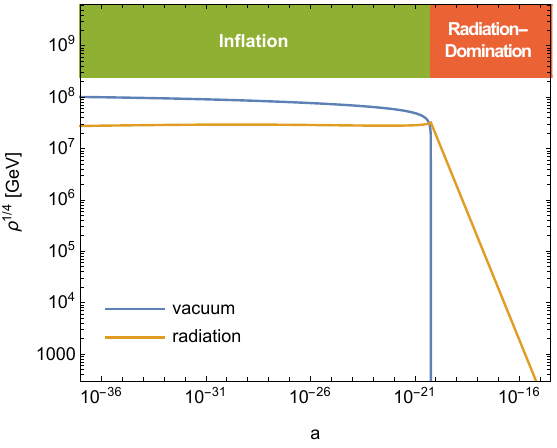}
    \caption{Evolution of the vacuum and radiation energy densities in chain inflation scenarios with and without a post-inflationary phase transition. In the left panel one slower phase transition occurs after inflation (boundary condition (i)). The Universe becomes vacuum-dominated for a second (short) period before the last transition transfers the remaining vacuum energy into radiation. In the right panel inflation ends by the last phase transition and the Universe immediately enters the standard radiation-dominated epoch (boundary condition (ii)). In the above figures the duration of the phase transitions has been neglected. The following parameters were chosen: $V_*^{1/4}=10^8\:\text{GeV}$, $\Delta V^{1/4}=2.7\times 10^6\:\text{GeV}$, $\Gamma^{1/4}_*/H_*=3.0\times 10^4$, $S_1=-2$, $S_2=4$, $S_3=3$, $S_4=1$ and $\alpha=10$ ($\alpha=0$) in the left (right) panel.}
    \label{fig:densities}
\end{figure}

In Fig.~\ref{fig:densities} we depict the evolution of the vacuum and radiation energy density in chain inflation as a function of the scale factor. The left panel shows the case with a (slower) post-inflationary transition during an extended graceful exit. The right panel displays the case in which all vacuum energy has been released at the end of inflation (prompt graceful exit at the end of inflation). In both cases the vacuum energy is continuously converted into radiation energy during inflation. The corresponding increase of $\rho_{\text{rad}}$ is, however, compensated by the reduction of $\rho_{\text{rad}}$ due to redshifting such that, overall, $\rho_{\text{rad}}$ remains approximately constant during inflation. But once all vacuum energy is released, the radiation source drys up, and the existing radiation density redshifts as $\rho_{\text{rad}}\propto 1/a^4$ with the scale factor of the Universe until the present time. In the right panel this crossover occurs immediately at the end of inflation, while in the left panel a slower post-inflationary phase transition occurs which is preceded by a second short period of vacuum domination (at $a\sim 8\times 10^{-20}$ in the figure). Let us also note that, while Fig.~\ref{fig:densities} was obtained for a concrete parameter choice (see figure caption), the described features are generic to the entire chain inflation parameter space.

\section{CMB Constraints on Chain Inflation and Detection Prospects}

In order to constrain the effective theory of chain inflation and to assess future prospects to test its predictions, we need to determine the CMB observables and compare them to the existing data/ projected sensitivities of CMB experiments. In particular, we will employ the Planck observed normalization and spectral index of the scalar power spectrum to reduce the chain inflation parameter space. 

An important test of chain inflation models is given by the spectral index and its running or higher-order running measured by CMB experiments.  The simplest model of chain inflation, a pure titled cosine, would produce a single power-law form of the density perturbatsions.  However, in more general models,  chain inflation predicts a larger deviation of $\Delta_{\mathcal{R}}^2(k)$ from the power-law form (compared to slow-roll inflation).  Thus there also exist intriguing prospects to discover running or higher running of the spectral index with future ground- and space-based CMB experiments. The latter will, in particular, improve the sensitivity at small angular scales. We will investigate the prospects of future experiments to measure the scale-dependence of the scalar power spectrum produced by chain inflation.

\subsection{Present and Future CMB Measurements}\label{sec:measurement}

Let us first point out the relevant CMB measurements. Of particular importance is the Planck-observed normalization of the scalar power spectrum~\cite{Planck:2018jri} at the pivot scale,
\begin{equation}
    A_s=(2.10\pm 0.03)\times 10^{-9}\,,
\end{equation}
and the spectral index at the pivot scale\footnote{Here we quote the best-fit spectral index from~\cite{Planck:2018jri} in the presence of running and running-of-running.}~\cite{Planck:2018jri},
\begin{equation}
    n_s=0.963\pm 0.005\,.
\end{equation}
As we will show in the next section, the above two constraints essentially fix two parameters in the effective theory of chain inflation (see also Sec.~\ref{sec:cmb}).

Furthermore, Planck has provided bounds on the running and running-of-running of the scalar spectral index at the pivot scale~\cite{Planck:2018jri},
\begin{equation}\label{eq:dnsdata}
    d\log n_s/d\log k=0.002\pm 0.010\,,\qquad d^2\log n_s/d\log k^2= 0.010 \pm 0.013\,.
\end{equation}
We emphasize, however, that these bounds were obtained under the assumption of vanishing running-of-running-of-running and higher running. In contrast, chain inflation generically features non-vanishing $d^3\log n_s/d\log k^3$, $d^4\log n_s/d\log k^4$, $\dots$ which impact the shape of the power spectrum. Therefore, applying Eq.~\eqref{eq:dnsdata} to chain inflation would introduce unwanted inaccuracies.

Luckily, Planck has also performed a model-independent reconstruction of the scalar power spectrum as a function of the comoving scale in~\cite{Planck:2018jri}. The $2\sigma$-contour of the reconstructed power spectrum is depicted as the gray band in Fig.~\ref{fig:cmbspectra}. For the purpose of constraining the scale dependence of $\Delta_{\mathcal{R}}^2(k)$ in chain inflation, we can thus directly compare the predicted power spectrum in the CMB range with the reconstructed power spectrum.

Future CMB observatories will substantially improve the power spectrum measurement at small angular scales (large $k\simeq 0.1-1\:\text{Mpc}^{-1}$) compared to Planck. In order to asses future prospects to detect a non-vanishing running (or higher running) of the spectral index produced by chain inflation, we will consider the forecasted sensitivity of the Simons Observatory to the scalar power spectrum.\footnote{The sensitivity forecast is presented in Fig.~21 of~\cite{SimonsObservatory:2018koc} with the uncertainty on the optical depth $\tau$ factored out which we assume to be similar as for Planck.}~This sensitivity -- which we depict as the pink band in Fig.~\ref{fig:cmbspectra} -- is presented in~\cite{SimonsObservatory:2018koc} in the form of a projected uncertainty band on $\Delta_{\mathcal{R}}^2(k)$ assuming a baseline $\Lambda$CDM model without running of the spectral index. In the following we will make the simplified assumption that Simons Observatory will be able to detect a deviation of $\Delta_{\mathcal{R}}^2(k)$ from the power-law form whenever the predicted spectrum falls outside Simons Observatory's projected $\Lambda$CDM uncertainty band.

The scalar power spectrum at even smaller angular scale (larger $k$) can be accessed through its impact on spectral distortions in the CMB. If funded, future proposed satellite experiments such as PIXIE~\cite{Kogut:2011xw} will potentially probe normalizations down to $\Delta_{\mathcal{R}}^2\sim 10^{-8}$ at $k=(10^2-10^4)\:\text{Mpc}^{-1}$. While PIXIE's projected sensitivity would  still be too low to probe a power-law-type scalar power spectrum, it could be sensitive to inflation models with a positive running (or higher running) of the spectral index. In order to estimate its discovery potential, we will employ PIXIE's power spectrum sensitivity curve provided in~\cite{Byrnes:2018txb} (orange contour in Fig.~\ref{fig:cmbspectra}). We will show that PIXIE would be able to probe a sizeable fraction of the chain inflation parameter space.

\subsection{Fixing Chain Inflation by the CMB Amplitude and the Spectral Index}
\label{sec:fixingparameters}

In order to constrain chain inflation we need to determine the CMB observables. In the first step we will derive the CMB normalization and the scalar spectral index within the effective theory of chain inflation (which captures a large fraction of possible models). As noted in Sec.~\ref{sec:parameterization} a parameter point is characterized by the choice of eight input parameters  $\{V_*,\, \Delta V,\, \Gamma_*,\, S_1,\, S_2,\, S_3,\, S_4,\, \alpha\}$, where the first seven determine the dynamics of chain inflation, while $\alpha$ characterizes the graceful exit from chain inflation. The special case of zero post-inflationary phase transitions corresponds to $\alpha=0$, otherwise $\alpha > 0$. 

 The normalization of the scalar power spectrum at the Pivot scale follows from Eq.~\eqref{eq:powerspectrum},
\begin{equation}\label{eq:Asmodel}
    A_s=0.06 \left(\frac{\Gamma_*^{1/4}}{H_*}\right)^{-5/3}\,,
\end{equation}
where $H_*$ can be expressed in terms of the input parameters through Eq.~\eqref{eq:rhoradstar}. Furthermore, the spectral index at the pivot scale is obtained by plugging Eq.~\eqref{eq:Gammaparametrization} into Eq.~\eqref{eq:ns},
\begin{equation}\label{eq:nsmodel}
    n_s \simeq 1 -\frac{5}{12}\frac{\Delta V}{V_*}\frac{1.4\,\Gamma_*^{1/4}}{H_*} \left( 2- S_1     \right)\,.
\end{equation}

In order to study the phenomenology of viable chain inflation realizations, we want to limit our analysis to the parameter space which is consistent with present CMB bounds. In the following we will, therefore, take $V_*$, $S_{2,3,4}$ and -- in the case of a post-inflationary transitions -- $\alpha$ as input parameters. The three remaining parameters $\Gamma_*$, $\Delta V$ and $S_1$ will be fixed by  by imposing the observational constraints on $A_s$ and $n_s$ as well as by requiring the correct number of e-folds during (observable) inflation. Specifically, we will fix
\begin{itemize}
    \item $\Gamma_*$ through Eq.~\eqref{eq:Asmodel} by imposing the CMB normalization $A_s=2.10\times 10^{-9}$~\cite{Planck:2018jri},
    \item $S_1$ through Eq.~\eqref{eq:nsmodel} by requiring the spectral index $n_s=0.963$~\cite{Planck:2018jri},
    \item $\Delta V$ (or equivalently the total number of phase transitions) through Eq.~\eqref{eq:astar} by imposing the scale factor $a_0=1$ of today's Universe,
\end{itemize}
where we neglected the small experimental uncertainty on $A_s$ and $n_s$ (which does virtually not affect the predictions of this work). Let us also note that the parameter determination is done iteratively, i.e.\ we vary these parameters and calculate the observables until the three above-listed equations are fulfilled.

The remaining parameter space spanned by $\{V_*,\, S_2,\, S_3,\, S_4,\, \alpha \}$ can be constrained through the running and higher running of the spectral index as we will describe in the next section.

\subsection{Deviation of the Scalar Power Spectrum from the Power-Law Form}\label{sec:scaledependence}

Apart from the spectral index, the scale-dependence of the scalar power spectrum is typically characterized by the running and the running-of-running of the spectral index. By applying Eq.~\eqref{eq:runningns} to the effective theory of chain inflation we find,
\begin{align}\label{eq:runningns2}
\frac{d\log n_s}{d\log k} &= \frac{5}{12}\left(\frac{\Delta V}{V_*}\frac{1.4\Gamma_*^{1/4}}{H_*}\right)^2
\left(\frac{-S_1^2}{4}+2\,S_2 + S_1 -3\right)\,,\nonumber\\
\frac{d^2\log n_s}{d\log k^2} &=
\frac{5}{12}\left(\frac{\Delta V}{V_*}\frac{1.4\Gamma_*^{1/4}}{H_*}\right)^3
\left(
\frac{S_1^3}{8}-2 \,S_1 S_2 + 6\, S_3 - \frac{3\,S_1^2-16\,S_2}{4}
+\frac{7\,S_1}{2} -9
\right)\,.
\end{align}

While the above expressions can be useful to quickly assess the scale-dependence of the scalar power spectrum for a given parameter combination, it is preferable to calculate the full power spectrum as a function of the comoving scale. This is because experimental constraints on $d\log n_s/d\log k$ and $d^2\log n_s/d\log k^2$ have only been published for the special case in which $d^3\log n_s/d\log k^3$ (and higher running) vanishes -- which is not the case in chain inflation (see Sec.~\ref{sec:measurement}). The full power spectrum, on the other hand, can directly be compared to the Planck-reconstructed power spectrum as well as to the projected sensitivities of the Simons Observatory and PIXIE.

For the purpose of calculating $\Delta_{\mathcal{R}}^2(k)$ we follow the approach outlined in Sec.~\ref{sec:evolution}. In the first step we derive the evolution of the energy densities $\rho_{\text{vac}}(\tilde{\phi})$, $\rho_{\text{rad}}(\tilde{\phi})$ by solving Eq.~\eqref{eq:rhodgl2} numerically. This also allows us to determine $H(\tilde{\phi})$ through Eq.~\eqref{eq:Hubble}. Plugging the so-obtained $H(\tilde{\phi})$ as well as $\Gamma(\tilde{\phi})$ from Eq.~\eqref{eq:Gammaparametrization} into Eq.~\eqref{eq:powerspectrum} then yields $\Delta_{\mathcal{R}}^2(\tilde{\phi})$. Finally, in order to express the power spectrum in terms of the comoving scale $k$, we employ Eq.~\eqref{eq:matching}, where we set $a_* = k_*/H_* $ with $k_*=0.05\:\text{Mpc}^{-1}$ (see Sec.~\ref{sec:evolution}).

\begin{figure}[htp]
    \centering
    \includegraphics[width=0.7\textwidth]{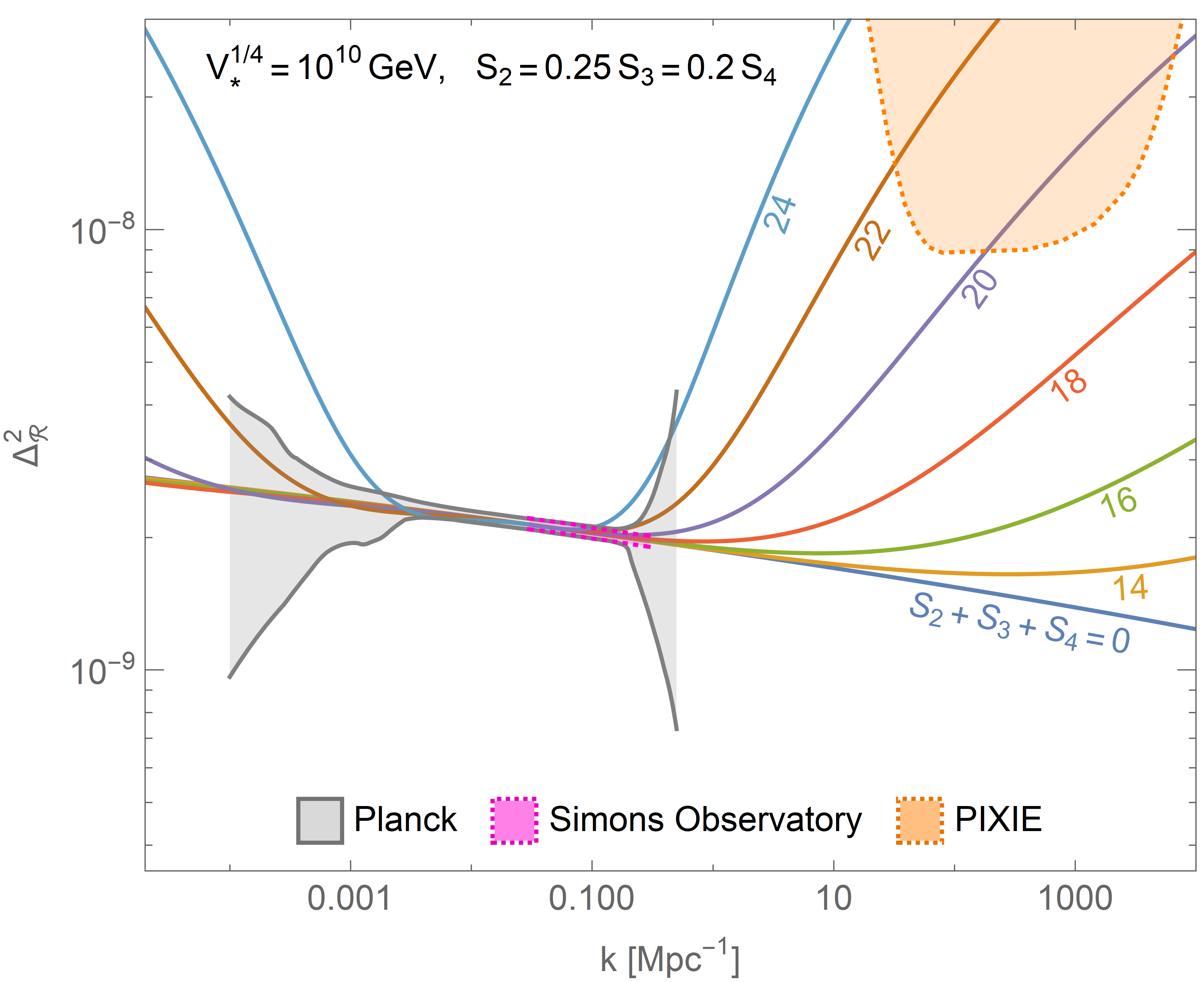}
    \caption{Scalar power spectrum of chain inflation for the parameter choices indicated in the figure (the numbers below the curves refer to the value of $S_2+S_3+S_4$). Curves which fall inside the gray region are compatible with Planck. Curves outside (inside) the pink band are expected to produce a signal (no signal) of a running spectral index at the Simons Observatory. Power spectra crossing the orange region cause a CMB spectral distortion which could be measured by proposed satellite PIXIE.}
    \label{fig:cmbspectra}
\end{figure}

The scale-dependence of the power spectrum in chain inflation originates (mostly) from the variation of the tunneling rate along the chain of vacua which is caused by the parameters $S_{1,2,3,4}$ (cf.~Eq.~\eqref{eq:Gammaparametrization}). While $S_1$ mostly affects the spectral index, $S_{2,3,4}$ induce running and higher running of the spectral index (cf. Eq.~\eqref{eq:nsmodel} and Eq.~\eqref{eq:runningns2}). The larger $S_{2,3,4}$, the more $\Delta_{\mathcal{R}}^2(k)$ deviates from the power-law form. This is illustrated in Fig.~\ref{fig:cmbspectra}, where we depict $\Delta_{\mathcal{R}}^2(k)$ for different choices of $S_2+S_3+S_4$ (assuming a fixed ratio $S_2=0.25 \,S_3 = 0.2\, S_4$, an inflation scale $V_*^{1/4}=10^{10}\:\text{GeV}$ and $\alpha=0$). Also shown is the Planck reconstructed power spectrum ($2\sigma$-contour), the projected power spectrum reconstruction by the Simons Observatory and the sensitivity projection for PIXIE (see Sec.~\ref{sec:measurement} for details). It can be seen that $S_2+S_3+S_4<22$ 
is required by the Planck data in this example. For $S_2+S_3+S_4=19-22$ the predicted scale-dependence of $\Delta_{\mathcal{R}}^2(k)$ is consistent with Planck, but testable by the Simons Observatory (about to start taking data) and in principle by future satellites that have been proposed (e.g. PIXIE). 
We note that the exclusions and future sensitivities somewhat depend on the ratio of coefficients $S_{2,3,4}$ and the scale of inflation (while $\alpha$ has a minor impact on the power spectrum). But in general, regions with $S_2+S_3+S_4\gg 10$ tend to produce a deviation from the power-law form which falls into the sensitivity window of CMB experiments. We comment that values of the coefficients $S_i=\mathcal{O}(10)$ are fully realistic in chain inflation models. For instance, if we consider the Lagrangian in Eq.~\eqref{eq:tiltedcosine}, such values of the $S_i$ merely correspond to $\mathcal{O}(1\%)$ changes of the parameters $f,\mu,\Lambda$ over the course of inflation. Notice also that the CMB constraints in Fig.~\ref{fig:cmbspectra} effectively impose $|S_i|<S_{E,*}$ (once we take into account that the bounce action at horizon crossing of the CMB scales must fall in the range of $S_{E,*}=30-120$ in order to obtain the correct CMB normalization). Hence, they provide a strong justification for our effective treatment of chain inflation in terms of the expansion in Eq.~\eqref{eq:Sexpansion}.

\section{Searching for Gravitational Waves from Chain Inflation}\label{sec:SearchingGW}

In slow-roll inflation observable gravitational waves may originate from the primordial tensor fluctuations of the metric. Such inflationary gravitational waves can be detected indirectly through a polarization signal in the CMB. Chain inflation, in contrast, is not expected to yield a polarization signal. This is because the magnitude of the tensor modes is set by the scale of inflation which is suppressed (compared to the Planck scale) in chain inflation~\cite{Freese:2021noj} (cf.\ Eq.~\eqref{eq:maxscale}).

Nevertheless, it turns out that chain inflation, produces a striking gravitational wave signal which stems from the bubble collision and bubble expansion during the first-order phase transitions. This gravitational wave signal does not fall into the frequency band of CMB experiments. Rather, it can directly be detected by pulsar timing arrays or interferometer experiments. 

\subsection{Origin of the Gravitational Wave Signal}\label{sec:originGW}

In this section we will describe the derivation of the gravitational wave spectrum in chain inflation. First-order phase transitions -- as they occur during chain inflation -- can be the source of strong gravitational radiation~\cite{Witten:1984rs,Hogan:1986qda}. The latter is generated by the collisions of the vacuum bubbles which form and expand during the phase transitions~\cite{Kosowsky:1992rz,Kosowsky:1992vn}. Furthermore, gravitational waves can be induced by sound waves~\cite{Hindmarsh:2013xza,Hindmarsh:2015qta,Hindmarsh:2017gnf} and magneto-hydrodynamic turbulence~\cite{Kosowsky:2001xp,Dolgov:2002ra,Caprini:2009yp} which form in the surrounding radiation plasma when the bubbles expand.
The relative importance of the different contributions depends on the underlying microphysics. If the energy density of the preexisting radiation plasma is small compared to the released vacuum energy (strong first-order phase transition) and/or if the coupling between the tunneling scalar field and the radiation degrees of freedom is sufficiently suppressed, the bubbles tend to reach a ``runaway regime'', in which they expand practically unhindered and all the energy of the phase transition is released during the bubble collisions. In this case the gravitational wave signal originates dominantly from the bubble collisions. In the opposite regime in which the preexisting plasma density is larger or comparable to the vacuum energy released in the phase transition, and in which unsuppressed couplings between the tunneling field and the radiation exist, the plasma inflicts a substantial pressure on the expanding bubbles. The bubble walls tend to reach a constant velocity quickly and lose most of their energy by heating and creating bulk motion in the plasma (see e.g.~\cite{Espinosa:2010hh}). In this case, the sound waves and magneto-hydrodynamic turbulence are the main sources of gravitational waves. Realistic chain inflation models may fall into either of the described categories (bubble-wall-dominated or a plasma-dominated signal) or may give rise to comparable signal contributions from both sources. It may also occur that the relative importance of the two contributions changes over the course of chain inflation. In order to keep the discussion simple, while capturing the range of possible signals, we will focus on the two limiting cases in which 
the signal either comes entirely from the bubble collisions or entirely from sound waves. We note that we will always neglect the magnetohydrodynamics contribution which can potentially be comparable to the sound wave contribution, but suffers from large theoretical uncertainties as well as from a strong model-dependence (see e.g.~\cite{Caprini:2018mtu}).

Since each phase transition during chain inflation completes before the start of the next phase transition, we treat them independently. In particular, we determine the gravitational wave spectrum by taking the superposition of the contributions from the individual transitions. Expressing the gravitational wave energy density (as a function of frequency) in terms of the critical density we thus have,
\begin{equation}\label{eq:sumspectrum}
    \Omega_{\text{GW}}(f) = \sum\limits_i  \Omega_{\text{GW},i}(f) \,,
\end{equation}
where the sum runs over all phase transitions which occur either during inflation or during the graceful exit (i.e.\ including the post-inflationary transition if there is one). We note that this picture may be overly simplistic for the acoustic plasma-induced gravitational waves because, for instance, sound waves
continue to act as a source of gravitational radiation well after the merging of the bubbles has completed. Therefore, collective effects on the bulk motion could occur and one could even hope for resonant enhancements of the sound waves induced by subsequent transitions. We leave a more detailed investigation of collective effects -- which will require dedicated hydrodynamic simulations -- for future work and, in the following, simply employ Eq.~\eqref{eq:sumspectrum} to obtain the gravitational wave spectrum. 

\paragraph*{Gravitational Waves from Bubble Collisions\\}
Let us first focus on the runaway-regime where the gravitational wave signal is dominated by the bubble collisions. This contribution is most commonly estimated in the so-called ``envelope approximation'' which places the stress-energy in a thin shell at the bubble wall which disappears upon collision~\cite{Kosowsky:1992rz,Kosowsky:1992vn,Huber:2008hg}. 
Hence, only the uncollided envelope of the spherical bubbles is taken into account in calculating the gravitational wave production, while the shear stress after the collision is neglected.
The envelope approximation applies to phase transitions in the thin-wall regime of vacuum tunneling in which the tunneling field becomes trapped temporarily in the old vacuum within the bubble collision region~\cite{Hawking:1982ga,Watkins:1991zt,Falkowski:2012fb} (minimizing the shear stress after collision)~\cite{Jinno:2019bxw}. However, in the opposite thick-wall regime, the tunneling field does not get trapped and rather undergoes oscillations around the new vacuum within the bubble overlap region. This leads to substantial propagation of the shear stress after collision in conflict with the assumption of the envelope approximation~\cite{Cutting:2020nla}. Simulations dedicated to the thick-wall regime have, for instance, been performed in~\cite{Cutting:2020nla}\footnote{The thick-wall case corresponds to the smallest $\bar{\lambda}$ simulated in~\cite{Cutting:2020nla}} (denoted as ``thick-wall simulation'' in the following). The predicted gravitational wave spectrum significantly differs between the envelope approximation and the thick-wall simulation. Differences arise in particular in the infrared (ultraviolet) tail of the spectrum which is harder (softer) in the envelope approximation compared to the thick-wall simulation. Similar deviations from the envelope approximation have also been observed in simulations of strong phase transitions associated with the breaking of a gauge symmetry~\cite{Lewicki:2020jiv,Lewicki:2020azd}. In order to reflect these theoretical uncertainties in the produced spectrum, we will present the gravitational wave signal from chain inflation separately for the envelope approximation and the thick-wall simulation.

\begin{table}[t]
\begin{center}
\begin{tabular}{|cccc|}
\hline
&&&\\[-4mm]
 & $\quad\widetilde{\Omega}\quad$ & $\quad b\quad$ & $\quad c \quad$ \\
 \hline
&&&\\[-4mm] 
 envelope & $1$ & $2.8$ &  $1$ \\
 thick-wall & $0.35$ & $0.7$ &  $2.2$\\ \hline
\end{tabular}
\end{center}
\vspace{-0.4cm}
\caption{Parameters entering the gravitational wave spectrum from bubble collisions in a first-order phase transition (Eq.~\eqref{eq:gravityspectrum}) as determined in the envelope approximation (taken from~\cite{Huber:2008hg}) and in the thick-wall simulation~\cite{Cutting:2020nla}. 
}
\label{tab:gwparameters}
\end{table}

The contribution to the gravitational wave spectrum by the $i$-th phase transitions -- assuming that all energy is released by the bubble collisions (bc) -- takes a broken power-law form~\cite{Kosowsky:1992rz,Kosowsky:1992vn,Caprini:2018mtu}\footnote{Notice that our expression for $\Omega_{\text{GW}}^{\text{bc}}$ is more general than, for instance, Eq.~(340) in~\cite{Caprini:2018mtu}. This is because~\cite{Caprini:2018mtu} assumes a radiation-dominated Universe and no further entropy production after the phase transition. For this special case one can express $a_i^4=6\times 10^{-51}\:\text{GeV}^4/(\Delta V g_{\text{eff}}(T)^{1/3})$ in our Eq.~\eqref{eq:gravityspectrum} such that Eq.~(340) in~\cite{Caprini:2018mtu} is indeed recovered. However, this special case does not apply to the phase transitions during inflation, which is why we explicitly keep the scale-factor dependence in Eq.~\eqref{eq:gravityspectrum}.},
\begin{equation}\label{eq:gravityspectrum}
\Omega_{\text{GW},i}^{\text{bc}}(f)\: h^2  = 5.46\times 10^7\:\text{GeV}^{-6}\; \widetilde{\Omega}\; \left(\frac{\Delta V}{\beta_i}\right)^2\:a_i^4\times \frac{(b+c)\left(f/f_{\text{peak}, i}^0\right)^b}{c+b\left(f/f_{\text{peak}, i}^0\right)^{b+c}}
\,,
\end{equation}
where the normalization $\tilde{\Omega}$, infrared and ultraviolet power-law indices $b$ and $c$ are given in Tab.~\ref{tab:gwparameters} for the envelope approximation and the thick-wall simulation.\footnote{Notice a slightly different notation compared to our previous work~\cite{Freese:2022qrl}. The parameters $a$, $b$ of~\cite{Freese:2022qrl} are called $b$, $c$ in this work to avoid confusion with the scale factor of the universe. Furthermore, the parameter $\widetilde{\Omega}$ we employ in this work is different by a factor of $1/0.077$ compared to the $\widetilde{\Omega}$ defined in~\cite{Freese:2022qrl}.} The factor $a_i^4$ accounts for the redshift of the gravity wave amplitude from production until now ($a_i$ stands for the scale factor at the $i$-th phase transition which can be derived from Eq.~\eqref{eq:matching}). Furthermore, the redshifted peak frequency $f_{\text{peak},i}^0$ is given by,
\begin{equation}\label{eq:peakf_emission}
 f_{\text{peak},i}^0 = a_i \,f_{\text{peak},i} \simeq 0.2\, a_i\, \beta_i\,,
\end{equation}
where $f_{\text{peak},i}$ stands for the peak frequency at emission which is extracted from the simulations~\cite{Huber:2008hg,Cutting:2020nla}. Here and in Eq.~\eqref{eq:gravityspectrum} the parameter $\beta_i$ is the inverse of the time-duration of the phase transition. From the simulations of bubble collisions performed in~\cite{Winkler:2020ape} we extract,

\begin{equation}\label{eq:betai}
   \beta_i \simeq \begin{cases} 2.8 \,\Gamma_i^{1/4}\,\quad\qquad\qquad\qquad\qquad\text{(transitions during chain inflation)}\\ \beta(\alpha) \text{ from Fig.~1 in~\cite{Freese:2022qrl}}\,\qquad\text{(slower transition during graceful exit)} \end{cases}
\end{equation}
where $\Gamma_i$ is the tunneling rate of the $i$-th vacuum transition. The upper expression in Eq.~\eqref{eq:betai} is valid in the regime where the phase transition occurs within a small fraction of a Hubble time such that the expansion of the Universe during the phase transition can be neglected. This condition is satisfied for all phase transitions during chain inflation. However, if there occurs a post-inflationary phase transition during an extended graceful exit from chain inflation, this last phase transition is considerably slower, and the expansion of the Universe cannot be neglected for calculating its duration. The $\beta_i$ of this last phase transition is a function of $\alpha$ and can be extracted from Fig.~1 in a previous paper by two of us~\cite{Freese:2022qrl}.

An important remark is that gravitational waves produced early during inflation suffer from a much stronger redshift compared to those produced towards the end of chain inflation. Indeed we will find that the gravitational radiation is strongly dominated by those waves produced during the last e-fold of inflation and during the graceful exit from chain inflation. The signal from slower transition(s) during the graceful exit can be particularly strong due to the scaling $\Omega_{\text{GW},i}^{\text{bc}}\: h^2(f)\propto 1/\beta_i^2$.

\paragraph*{Acoustic Gravitational Waves\\}
We now turn to the opposite regime in which the gravitational radiation dominantly stems from sound waves in the plasma, while the contribution from bubble collisions is negligible\footnote{In this case, radiation has been produced by earlier bubble collisions (typically before the observable inflation). Once there is significant radiation around, the bubbles experience friction and lose most of their energy to the plasma before collision. Hence, reheating is effectively achieved by heating the previously produced plasma.}. As noted earlier, this situation occurs if the plasma inflicts substantial pressure on the bubble walls during their expansion such that most of the vacuum energy is transferred to heat and bulk motion of the plasma before the bubbles collide. 
Because only the bulk motion can create anisotropic stress and subsequently gravitational waves, while the thermal energy cannot, the produced signal strength depends on the distribution between the two energy forms. We, therefore, introduce the efficiency factor $\kappa_i$ (where the index $i$ again stands for the $i$-th transition) which determines the fraction of the vacuum energy liberated in the phase transition which goes into bulk motion of the plasma. While $\kappa_i$, in principle, depends on the underlying microscopic theory, it was shown in~\cite{Espinosa:2010hh} that it can be related to the final bubble wall velocity and the ratio of released vacuum energy to radiation energy (prior to the transition) $\Delta V /\rho_{\text{rad},i}$. For concreteness we will focus on the case, where the bubble walls reach a velocity close to the speed of light, for which the following estimate exists~\cite{Espinosa:2010hh},
\begin{equation}
    \kappa_i \simeq \frac{\Delta V}{0.73\,\rho_{\text{rad},i}+0.083\sqrt{\Delta V \,\rho_{\text{rad},i}}+\Delta V}\,.
\end{equation}
Employing the efficiency factor from above we can calculate the sound-wave (sw) induced gravitational wave spectrum of the $i$-th transition~\cite{Hindmarsh:2015qta,Caprini:2018mtu}\footnote{Again, note that the difference between our expression for $\Omega^{\rm sw}_{\rm GW}$ and Eq.~(343) in~\cite{Caprini:2018mtu} is due to the fact that, in the latter, radiation domination and no further entropy production after the phase transition are assumed.},
\begin{equation}\label{eq:gravityspectrumsound}
\Omega_{\text{GW},i}^{\text{sw}}(f)\: h^2  = 1.12\times 10^8\:\text{GeV}^{-6}\: \left(\frac{\kappa_i^2\,\Delta V^2}{\beta_i\,H_i}\right)\:a_i^4\times \left(\frac{f}{f_{\text{peak}, i}^0}\right)^3\left(\frac{7}{4+3(f/f_{\text{peak}, i}^0)^2}\right)^{7/2}
\,,
\end{equation}
where $H_i$ denotes the Hubble scale at the $i$-th transition.

As in the bubble collision case (studied in the previous section), the redshifting by the factor $a_i^4$ suppresses gravitational waves which are produced early during inflation. The signal is again dominated by the emission during the last e-fold of inflation and, in particular, during the graceful exit from inflation (if a post-inflationary transition occurs). Compared to the bubble collision case, the amplitude is reduced by the efficiency factor $\kappa_i^2$ (since the produced thermal energy does not contribute to the gravitational wave signal). However, the suppression is (partially) compensated because Eq.~\eqref{eq:gravityspectrumsound} contains an extra factor of $\beta_i/H_i$ compared to Eq.~\eqref{eq:gravityspectrum} which arises because the sound waves continue to source gravitational waves long after the bubbles have collided (for about a Hubble time~\cite{Hindmarsh:2013xza}).

\subsection{Experimental Data and Future Sensitivities}

Let us now turn to the experiments which can detect the gravitational wave signal from chain inflation. A particular intriguing possibility -- which we pointed out in~\cite{Freese:2022qrl} -- is that chain inflation could have produced the stochastic gravitational wave background (tentatively) discovered by NANOGrav~\cite{NANOGrav:2020bcs,NANOGrav:2023gor}. We will now present our statistical method to identify gravitational wave spectra consistent with the NANOGrav signal. Afterwards we will investigate the sensitivity of further ongoing and upcoming gravitational wave observatories, including PTAs, ground and space interferometers, which can potentially measure the gravitational waves from chain inflation.

\subsubsection{Signal for a Stochastic Gravitational Wave Background}\label{sec:signal}
The NANOGrav collaboration has announced the detection of a 
low-frequency stochastic process which affects pulsar timing residuals~\cite{NANOGrav:2020bcs,NANOGrav:2023gor}. This signal has been been confirmed by the PPTA~\cite{Goncharov:2021oub,Reardon:2023gzh}, EPTA~\cite{Chen:2021rqp,EPTA:2023fyk} and CPTA~\cite{Xu:2023wog} pulsar timing array experiments. Within its latest (15-yr) data release NANOGrav, furthermore, reported evidence for quadrupolar Hellings-Downs correlations~\cite{Hellings:1983fr} associated with gravitational wave sources~\cite{NANOGrav:2023gor,NANOGrav:2023hvm}. The PTA observations, hence, likely amount to the discovery of a stochastic gravitational wave background at frequencies $f\sim\text{yr}^{-1}$. Among the most plausible sources for such a background in the sensitivity window of pulsar timing arrays are mergers of super-massive black-hole binaries~\cite{Rajagopal:1994zj,Jaffe:2002rt,Wyithe:2002ep,Sesana:2008mz,Burke-Spolaor:2018bvk,Middleton:2020asl}, a cosmic-string network~\cite{Vilenkin:1981bx,Vachaspati:1984gt,Damour:2004kw,Siemens:2006yp,Olmez:2010bi,Ringeval:2017eww,Ellis:2020ena,Blasi:2020mfx,Buchmuller:2020lbh} and first-order phase transitions~\cite{Caprini:2010xv,Schwaller:2015tja,Kobakhidze:2017mru,Nakai:2020oit,Addazi:2020zcj,Ratzinger:2020koh,Brandenburg:2021tmp,NANOGrav:2021flc,Borah:2021ocu,DiBari:2021dri,Lewicki:2021xku,Ashoorioon:2022raz,Freese:2022qrl,Freese:2023fcr,Cruz:2023lnq} -- of which the latter are the subject of this study. 

In the following we want to investigate, whether the NANOGrav signal can be explained by the gravitational radiation produced during chain inflation and during the graceful exit. For this purpose we constructed a binned list of 15 NANOGrav data points (with error bars) 
from the periodogram for a free spectral process provided in Fig.~3 of~\cite{NANOGrav:2023hvm}. This was done by extracting the median and 1$\sigma$-uncertainties from the probability distributions shown in the figure.
We only included the 15 frequency bins with $f<1\:\text{yr}^{-1}$ due to the potential noise contamination affecting the high-frequency bins~\cite{NANOGrav:2023hvm}. The so-obtained NANOGrav data set is depicted in the lower right panel of Fig.~\ref{fig:nopostcritical} (blue error bars). In order to identify the parameter regions favored by NANOGrav, we then performed a $\Delta\chi^2$-test using a standard $\chi^2$-metric.

\begin{figure}[htp]
    \centering
    \includegraphics[width=0.5\textwidth]{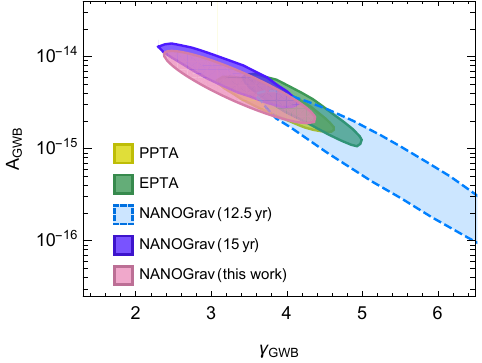}\vspace{-2mm}
    \caption{Amplitude and power law index consistent with the gravitational wave signal observed at several PTA experiments (2$\sigma$-contours). The published NANOGrav signal regions for the 15- and 12.5-yr data sets are shown in dark and light blue. The pink contour was derived with our $\Delta\chi^2$-test to the NANOGrav 15-yr data. The blue and yellow regions were obtained by the PPTA and EPTA experiments. Hence this figure shows that our statistical method upon the data accurately reproduces results previously published by the NANOGrav collaboration.}
    \label{fig:nanograv}
\end{figure}

In the first step we validated our statistical approach by deriving the NANOGrav signal region for pure power-law gravitational wave spectra. Following the NANOGrav conventions we defined the amplitude $A_{\text{GWB}}$ and power law $\gamma_{\text{GWB}}$ of a gravitational wave background which are related to the gravitational wave density (normalized to the critical density) in the following way~\cite{NANOGrav:2023hvm},
\begin{equation}\label{eq:powerlawspectrum}
    \Omega_{\text{GW}}(f)=\frac{2\pi^2 f^5}{3 H_0^2}A_{\text{GWB}}^2\left(\frac{f}{\text{yr}^{-1}}\right)^{-\gamma_{\text{GWB}}}\text{yr}^3\,,
\end{equation}
where $H_0$ denotes the Hubble constant. In Fig.~\ref{fig:nanograv} we compare the $2\sigma$-contour\footnote{For a power-law spectrum with free amplitude and free spectral index, the $2\sigma$-contour is defined by the condition $\Delta\chi^2\equiv\chi^2-\chi^2_{\text{min}}=6.2$, where $\chi^2_{\text{min}}$ is the $\chi^2$-value of the best-fit point.} in the $\gamma_{\text{GWB}}$-$A_{\text{GWB}}$-plane obtained from our $\Delta\chi^2$-test to the published NANOGrav (15-yr) $2\sigma$-contour~\cite{NANOGrav:2023gor} which was obtained with a complementary statistical approach. We also show the published signal region for the earlier (12.5-yr) NANOGrav data release~\cite{NANOGrav:2020bcs}, as well as for the PPTA~\cite{Reardon:2023gzh} and the EPTA experiments~\cite{EPTA:2023fyk}. It can be seen that our statistical method identifies a very similar parameter space as the official NANOGrav analysis and as the other PTA experiments. This proves that our statistical method is capable of identifying gravitational wave spectra consistent with the NANOGrav signal (and automatically also with the other PTA experiments). We can, hence, safely apply it to the more general gravitational wave spectra from chain inflation for which no official signal regions have been provided so far.

\subsubsection{Sensitivity Projections}\label{sec:sensitivity_projections}

While it is clearly exciting to consider chain inflation as the explanation of the NANOGrav signal, it may also turn out that the latter has a different physical origin -- for instance in the form of black hole mergers. In this light, it is crucial to also investigate complementary gravitational wave searches. In order to investigate the discovery potential for the gravitational waves from chain inflation we will compare the signal with the sensitivity of the relevant current and future observatories. The following experiments (going from low- to high-frequency sensitivities) will be considered:
\begin{itemize}
\item pulsar timing arrays: the (presently running) International Pulsar Timing Array experiment (IPTA)~\cite{Antoniadis:2022pcn} and the (planned) Square Kilometre Array Observatory (SKA)~\cite{Dewdney:2009}. Here, IPTA consists of the Nanohertz Observatory for Gravitational Waves (NANOGrav)~\cite{NANOGrav:2020bcs}, the European Pulsar Timing Array (EPTA)~\cite{Chen:2021rqp}, the Parkes Pulsar Timing Array (PPTA)~\cite{Goncharov:2021oub} and the Indian Pulsar Timing Array (InPTA)~\cite{Tarafdar:2022toa}.
\item space interferometers: the proposed future space interferometers Laser Interferometer Space Antenna (LISA)~\cite{LISA:2017pwj}, Big-Bang Observer (BBO)~\cite{Corbin:2005ny,Harry:2006fi}, Deci-Hertz Interferometer Gravitational-Wave Observatory (Decigo)~\cite{Seto:2001qf,Kawamura:2006up} and Ultimate-Decigo (U-Decigo)~\cite{Seto:2001qf,Kuroyanagi:2014qza}.
\item ground interferometers: the (operating) ground interferometer network consisting of the Advanced Laser Interferometer Gravitational-Wave Observatory (aLIGO)~\cite{Harry:2010zz,LIGOScientific:2014pky}, Advanced Virgo (aVirgo)~\cite{VIRGO:2014yos} and the Kamioka Gravitational-Wave Detector (KAGRA)~\cite{Somiya:2011np,Aso:2013eba} -- specifically we will consider the estimated aLIGO-aVirgo sensitivity of run O2 (LV$\,$O2) and the combined aLIGO-aVIRGO-KAGRA design sensitivity (LVK).  We will also consider the (upcoming) Einstein Telescope (ET)~\cite{Punturo:2010zz} and Cosmic Explorer (CE)~\cite{LIGOScientific:2016wof,Reitze:2019iox}.
\end{itemize}

For a given gravitational wave spectrum $\Omega_{\text{GW}}(f)$ (originating either from the bubble collisions or the sound waves created during chain inflation), we calculate the signal-to-noise ratio $\varrho$ for each experiment~\cite{Allen:1997ad,Maggiore:1999vm,Schmitz:2020syl},
\begin{equation}
    \varrho = \sqrt{n_{\text{det}}\, t_{\text{obs}}\int\limits_{f_{\text{min}}}^{f_{\text{max}}} df \left(\frac{\Omega_{\text{GW}}(f)}{\Omega_{\text{noise}}(f)}\right)^2}\,,
\end{equation}
and assume a signal is detectable for $\varrho>1$ following~\cite{Schmitz:2020syl}. In the above equation $t_{\text{obs}}$ stands for the observation time which we fix to $4\:$years for interferometers and to $20\:$years for projected sensitivities for pulsar timing arrays IPTA and SKA.\footnote{The $20\:$years observation time are only used to calculate the projected sensitivities of IPTA and SKA. In contrast for fitting the NANOGrav signal we use the actual observation time of 12.5-15 years as explained in Sec.~\ref{sec:signal}.} The parameter $n_{\text{det}}$ takes the value of one (two) for experiments which aim at detecting the gravitational wave signal via an auto-correlation (cross-correlation) measurement. Finally, the noise spectra of all considered experiments except U-Decigo are extracted from~\cite{Schmitz:2020syl}. The U-Decigo noise curve is estimated for the detector configuration suggested in~\cite{Kuroyanagi:2014qza} following the approach of~\cite{Ringwald:2020vei}. Note that we do not include potential astrophysical foregrounds in our sensitivity estimates which should thus be considered as optimistic.\footnote{For instance the sensitivity of the future space interferometers BBO, Decigo and U-Decigo at $f\lesssim 0.1\:\text{Hz}$ will potentially be affected by white dwarf confusion noise (see e.g.~\cite{Farmer:2003pa}). We note, however, that it is difficult to estimate the importance of this foreground since its magnitude is not yet known. Furthermore, one can hope to discriminate a cosmological gravitational wave background against the white dwarf foreground based on the spectral shape and through the resolution of individual white dwarfs.}

\subsection{Results: Gravitational Wave Signatures of Chain Inflation}

In the following we will present our results for the gravitational wave emission from chain inflation. We emphasize again that the gravitational waves from chain inflation are sourced by the first-order phase transitions (and not by the quantum fluctuations relevant in slow-roll inflation). We will first consider the case where chain inflation ends promptly with the last phase transition, before we turn to the case with a phase transition after inflation during an extended graceful exit. The signal from the graceful exit will turn out to be particular strong and to potentially explain the NANOGrav observations.

\subsubsection{Gravitational Waves Produced During Inflation}

Let us first assume that the Universe tunnels into its present vacuum at the end of chain inflation. No further transitions occur afterwards (within the lifetime of the Universe). The physics of such a prompt graceful exit has been described in Sec.~\ref{sec:exit}. All gravitational waves stem from the inflationary epoch in this case.

Within the effective theory of chain inflation (see Sec.~\ref{sec:parameterization}) the parameter space is spanned by the inflation scale $V_*^{1/4}$ and the three coefficients $S_{2,3,4}$ determining the time-evolution of the tunneling rate. The remaining parameters are fixed by the CMB normalization and by the spectral index (see Sec.~\ref{sec:fixingparameters}).

The gravitational radiation signal is mostly produced during the last $\mathcal{O}(1)$ e-fold of chain inflation. This is because the waves produced by earlier phase transitions suffer an enormous suppression by the redshifting. As a consequence, the gravitational wave amplitude is mostly sensitive to the sum $S_2+S_3+S_4$ -- which determines the tunneling rate towards the end of inflation\footnote{We remind the reader that the tunneling rate at the end of inflation is given by $\Gamma_c=\Gamma_*\,e^{-S_1-S_2-S_3-S_4}$. Since $S_1$ is fixed by requiring the observed spectral index, it is the choice of $S_2+S_3+S_4$ which mainly determines the strength of the gravitational wave signal from the end of chain inflation.} -- while the relative size of the $S_i$ plays only a subleading role. The larger $S_2+S_3+S_4$, the more the tunneling rate is reduced for the phase transitions at the end of chain inflation (which dominate the gravitational wave signal). Since slower transitions produce a stronger gravitational wave signal (cf. Eq.~\eqref{eq:gravityspectrum} and Eq.~\eqref{eq:gravityspectrumsound}) the gravitational wave amplitude grows with $S_2+S_3+S_4$.

In order to identify the chain inflation parameter space with an observable gravitational wave signal we performed scans over $V_*^{1/4}$ and $S_2+S_3+S_4$ (with the ratio $S_2=0.25 S_3 =0.2 S_4$ fixed for concreteness). Since inflation scales above $10^{12}\:\text{GeV}$ are excluded in chain inflation (at least for standard quasi-periodic potentials) we focused on the regime $V_*^{1/4}<10^{12}\:\text{GeV}$ (see Sec.~\ref{sec:cmb} and~\cite{Freese:2021noj}). We also note that the inflation scale cannot be arbitrarily low since the reheating temperature after the last transition should be above $1.8\:\text{MeV}$ for successful BBN. We will see that this leads to a lower limit on the scale of chain inflation $V_*^{1/4}\gtrsim 10\:\text{MeV}$.

Due to the theoretical uncertainties on the gravitational wave spectrum from first-order phase transitions we considered three distinct cases in which the emission during chain inflation is dominated by (1) the bubble collisions (modeled in the envelope approximation), (2) the bubble collisions (modeled in the thick-wall simulation) (3) sound waves in the plasma. We refer to Sec.~\ref{sec:originGW} for details.

\begin{figure}[t!]
\begin{center}
\includegraphics[width=0.45\textwidth,height=0.40\textwidth]{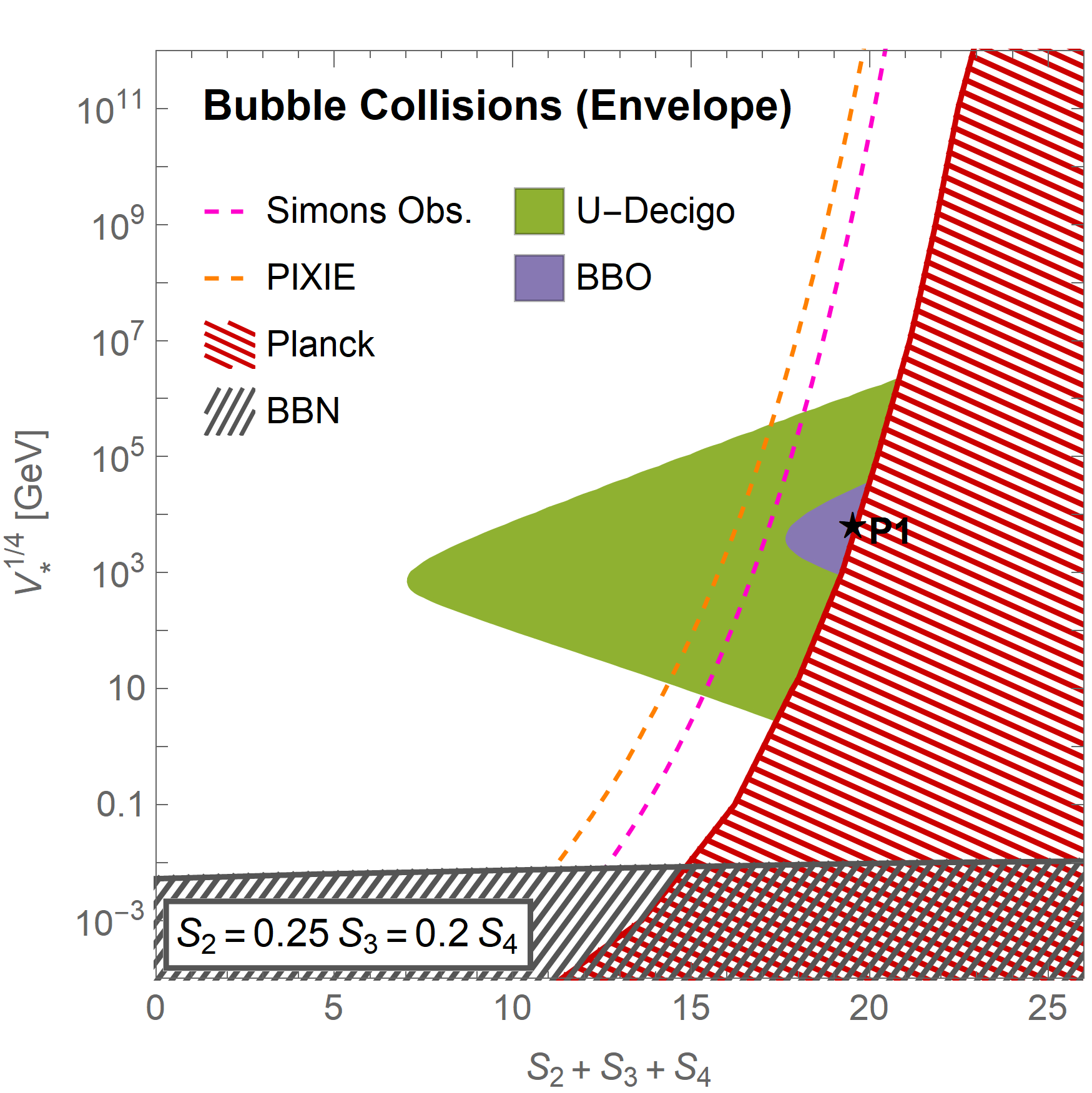}\hspace{2mm}
\includegraphics[width=0.45\textwidth,height=0.40\textwidth]{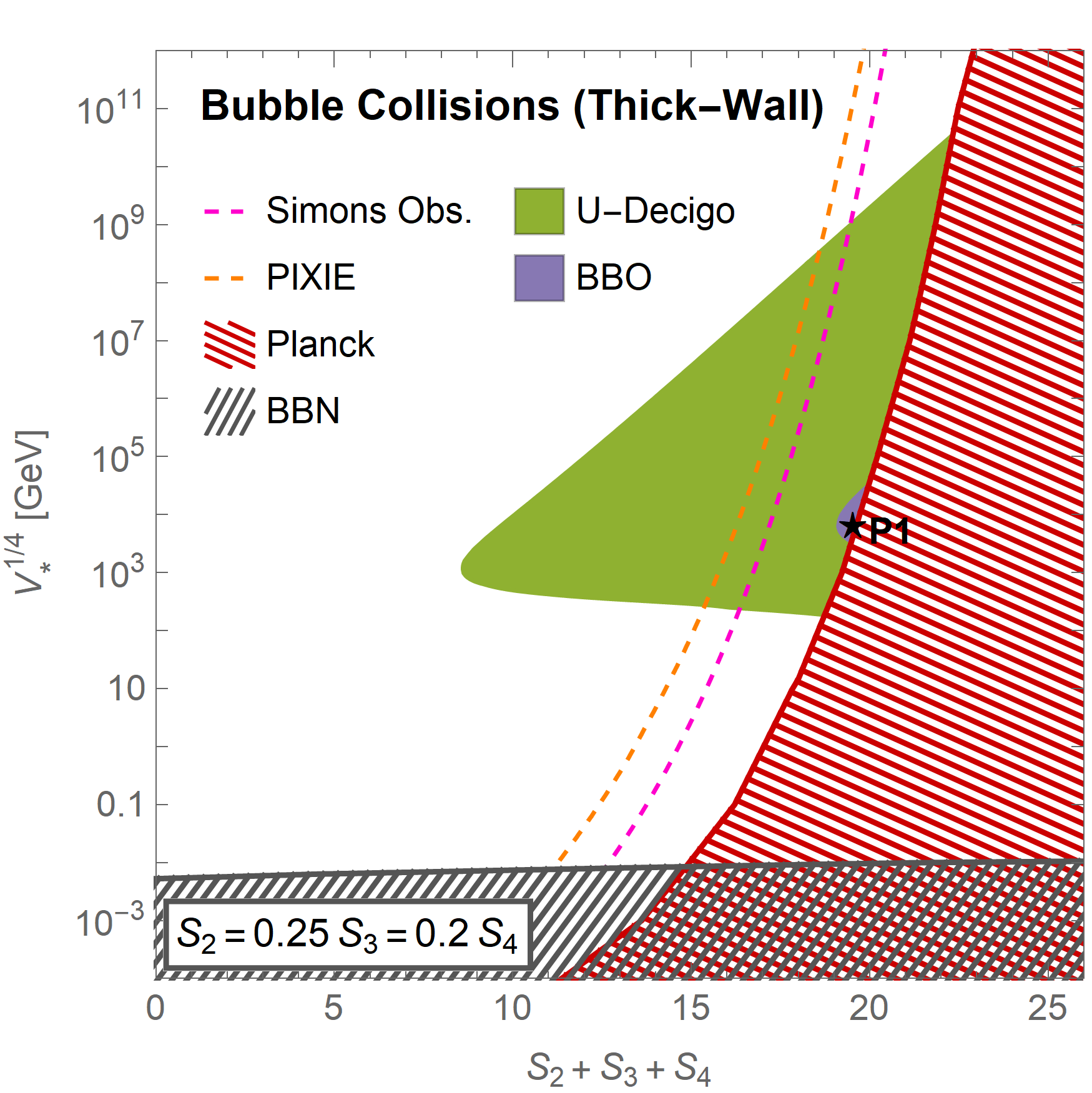}\\[1.mm]
\includegraphics[width=0.45\textwidth,height=0.40\textwidth]{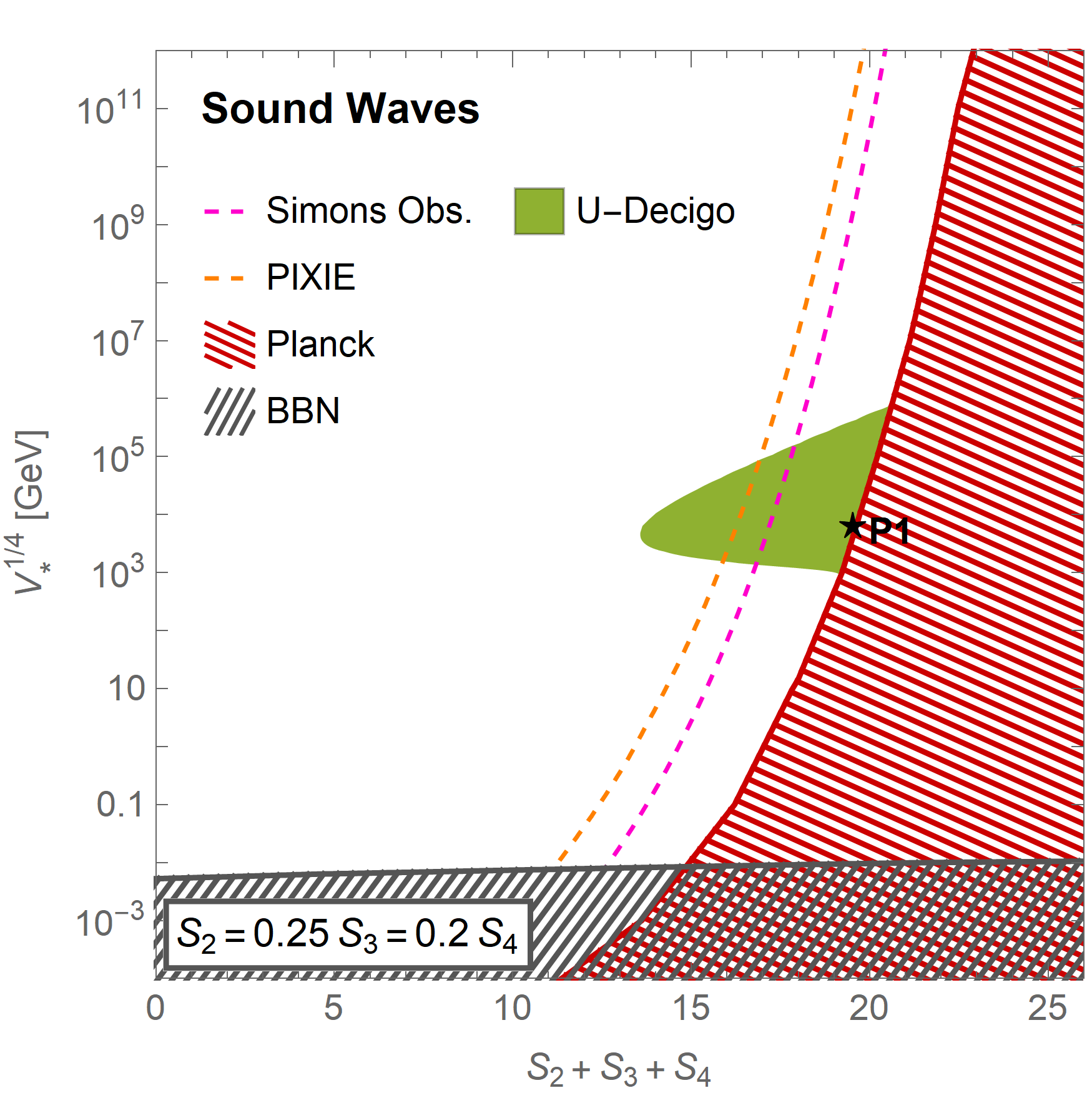}\hspace{2mm}
\includegraphics[width=0.45\textwidth,height=0.40\textwidth]{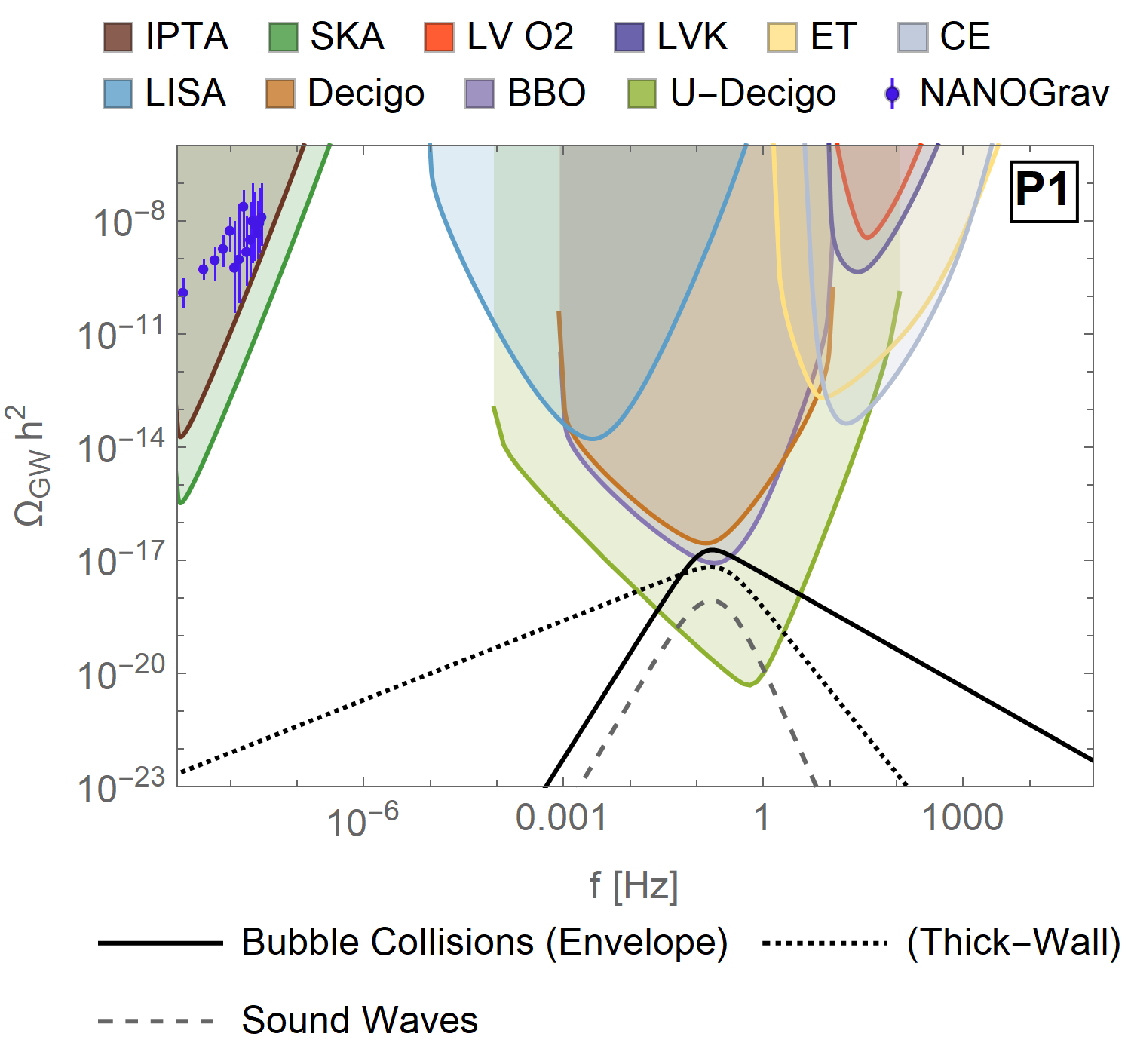}
\end{center}
\vspace{-5mm}
 \caption{Parameter scans and gravitational wave spectra in the effective theory of chain inflation with a prompt graceful exit (i.e.\ without phase transitions after inflation). Upper panels: Parameter scans assuming that the gravitational wave signal is dominated by bubble collisions.
 In the left (right) panel the envelope approximation (thick-wall simulation) is used to model the gravitational wave spectrum. In the color-shaded regions chain inflation induces a gravitational wave signal which can be probed by BBO and/or Ultimate-Decigo. In the black striped region the inflation scale is too low for BBN. The red striped region is excluded by Planck constraints on the scale-dependence of the scalar power spectrum. The regions right of the dashed pink (orange) line feature a large enough scale-dependence to be probed by complementary CMB measurements by the Simons Observatory (proposed PIXIE satellite). Lower left panel: same as the upper panels, but assuming that the gravitational wave emission dominantly stems from sound waves in the plasma (rather than from the bubble collisions). Lower right panel: gravitational wave spectra for the benchmark point P1 (see Tab.~\ref{tab:Benchmark points}) indicated in the parameter scans (for acronyms see Sec.~\ref{sec:sensitivity_projections}). Also shown are the NANOGrav data and the projected sensitivities (power-law-integrated sensitivity curves) of various ongoing and upcoming gravitational wave observatories.
}
    \label{fig:nopostcritical}
\end{figure}

The first three panels in Fig.~\ref{fig:nopostcritical} present the parameter regions which can be probed by gravitational wave experiments for the cases (1), (2), (3). Constraints from BBN and the CMB are also indicated. We find that only the proposed future space interferometers BBO and U-Decigo will reach the sensitivity to probe the gravitational waves radiated during chain inflation. While the gravitational wave amplitude grows with $S_2+S_3+S_4$, too large $S_2+S_3+S_4$ is excluded by Planck because it produces a too strong deviation of the scalar power spectrum from the power-law form (see Sec.~\ref{sec:scaledependence}). Within the entire Planck-allowed parameter space, the peak gravitational wave amplitude remains at $\Omega_{\text{GW}}h^2 \lesssim 10^{-17}$ -- requiring very advanced instruments to detect it. The best chances to measure the gravitational radiation from during chain inflation exist for an inflation scale $V_*^{1/4}\sim 10^4\:\text{GeV}$ for which the peak of the spectrum occurs at $f\sim \text{Hz}$, where space interferometers are most sensitive. This can be seen in the lower right panel of Fig.~\ref{fig:nopostcritical}, where we depict the gravitational wave spectrum for the benchmark point P1 (which features $S_2+S_3+S_4=19.6$, $V_*^{1/4}=6.3\:\text{TeV}$) together with the projected sensitivities of present and future instruments.

If BBO or U-Decigo report a gravitational wave signal in the future it will be intriguing to ask how one can identify chain inflation as its source. Assuming that the space interferometers can measure the peak of the spectrum (and not only the tail) the broken power-law shape characteristic for first-order phase transitions can be established. Since in chain inflation the spectrum is generated by the superposition of many phase transitions, a broader peak occurs compared to a single phase transition. Thus, the very specific spectral shape of its gravitational wave spectrum could potentially reveal chain inflation as the source of a future signal.

A strong indicator that chain inflation has been found in the data would be observations that confirm a correlation between the magnitude of the gravitational wave signal and the scale-dependence of the CMB scalar power spectrum. In Fig.~\ref{fig:nopostcritical} we also indicate the parameter ranges in which the  CMB experiments Simons Observatory and proposed PIXIE satellite will be able to measure a deviation of the scalar power spectrum from the power-law form.
As can be seen in Fig.~\ref{fig:nopostcritical} these parameter ranges strongly overlap with those, where a gravitational wave signal at BBO and/or U-Decigo is expected. Hence, we can hope for correlated gravitational wave and CMB signatures which would provide strong evidence for chain inflation.

\subsubsection{Gravitational Waves Produced by the Graceful Exit}

We now turn to the case with a phase transition after inflation during an extended graceful exit. As we argued in Sec.~\ref{sec:exit} such a slower, delayed phase transition is very common in chain inflation. This is because during chain inflation, the Universe populates highly unstable vacua, but today it has settled in a (meta)stable vacuum. In between, there likely occurs a vacuum with an intermediate lifetime whose decay marks the graceful exit from inflation.

Since the (delayed) graceful exit transition is slower compared to the transitions during inflation, and because it occurs at a lower redshift, it produces a gravitational wave signal which strongly dominates over the signal produced during chain inflation (cf. Eq.~\eqref{eq:gravityspectrum} and Eq.~\eqref{eq:gravityspectrumsound}).

In the effective theory of chain inflation, the magnitude and the peak frequency of the gravitational waves from the graceful exit are mostly determined by the strength of the phase transition $\alpha$ and the scale of inflation $V_*^{1/4}$. The $S_i$ -- while they have a major impact on the gravitational wave emission during chain inflation -- are less important for the graceful exit signal. 

\begin{figure}[htp]
\begin{center}
\includegraphics[width=0.45\textwidth,height=0.40\textwidth]{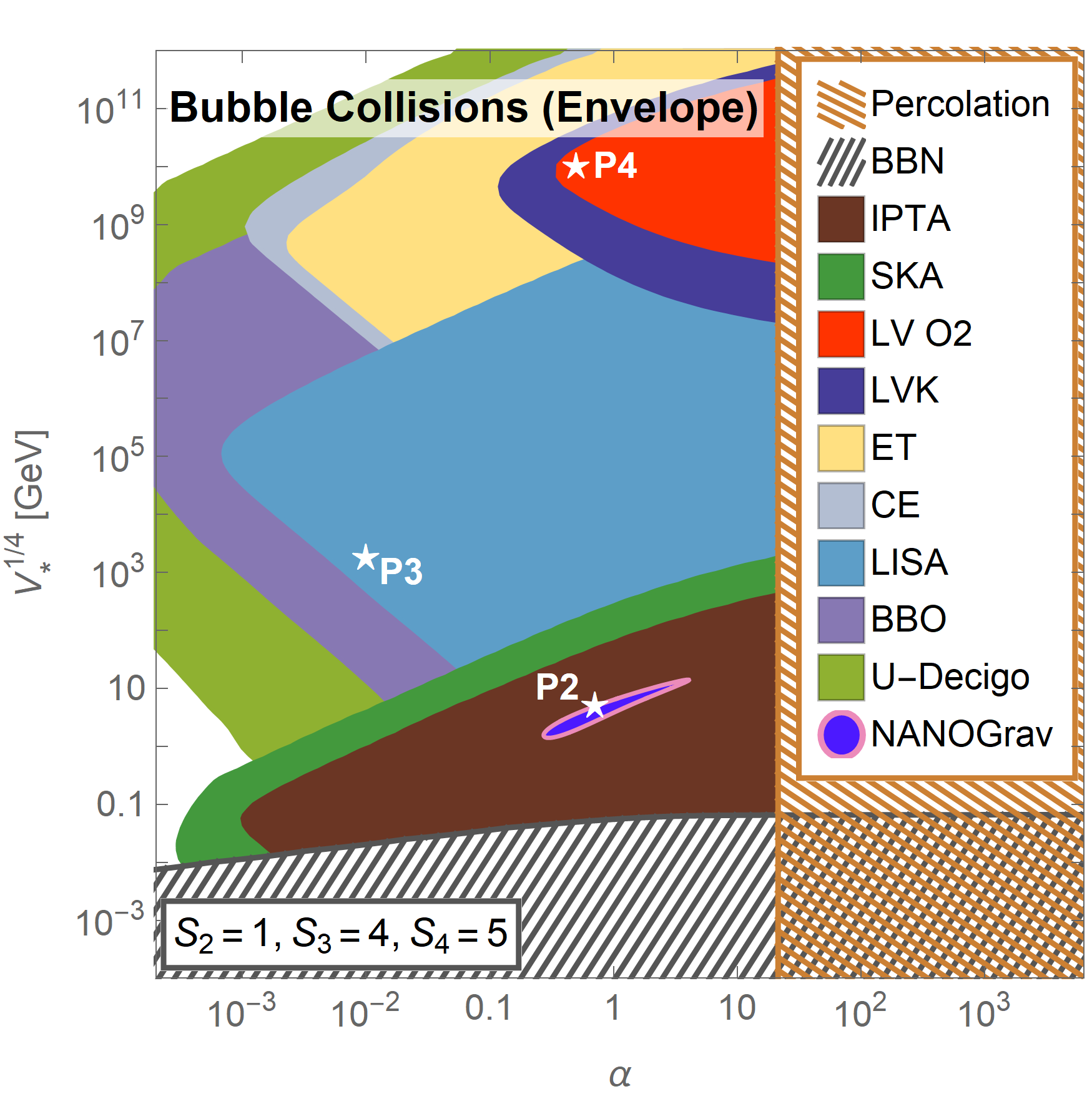}\hspace{2mm}
\includegraphics[width=0.45\textwidth,height=0.40\textwidth]{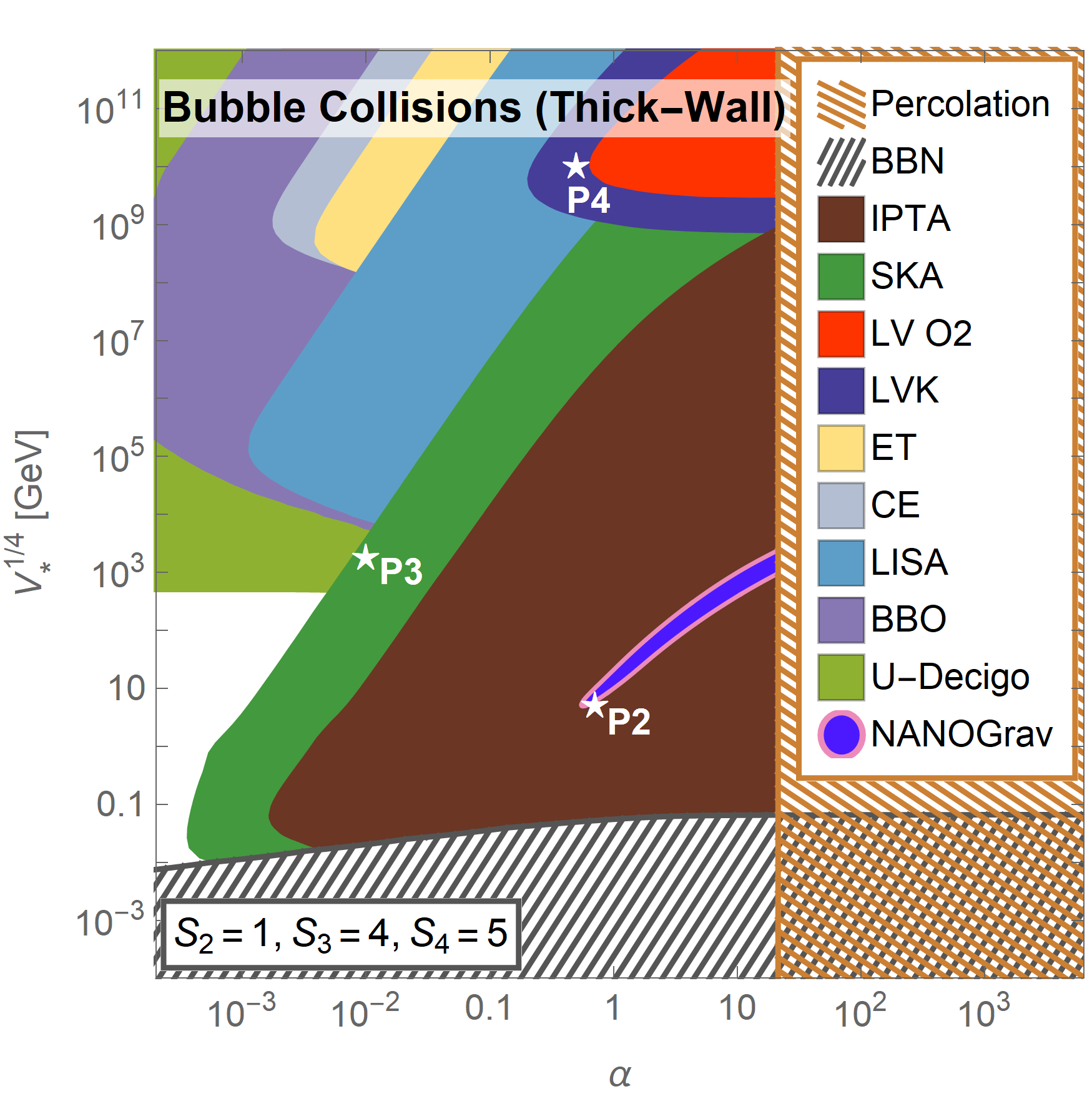}\\[1.mm]
\includegraphics[width=0.45\textwidth,height=0.40\textwidth]{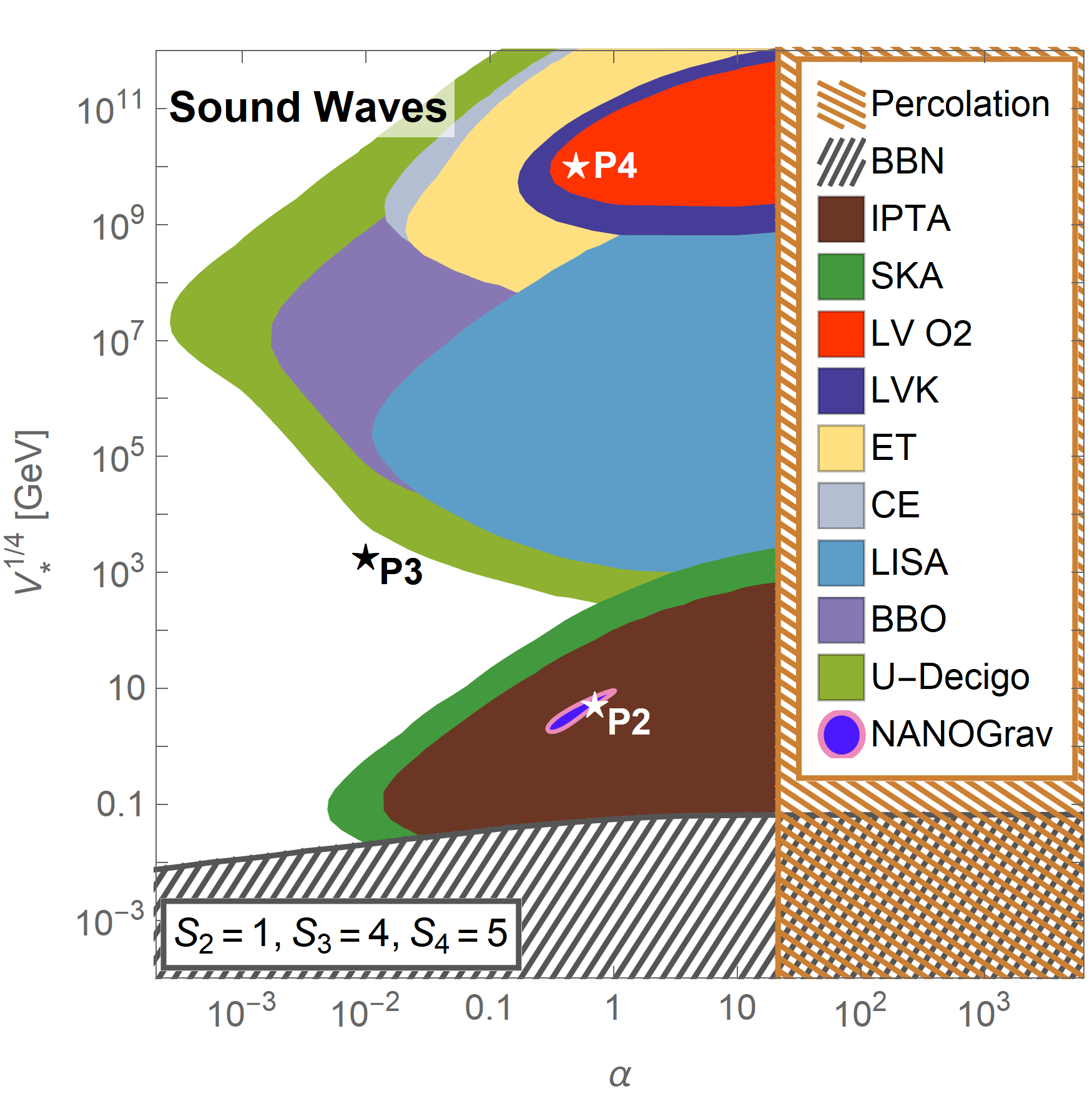}\hspace{2mm}
\includegraphics[width=0.45\textwidth,height=0.40\textwidth]{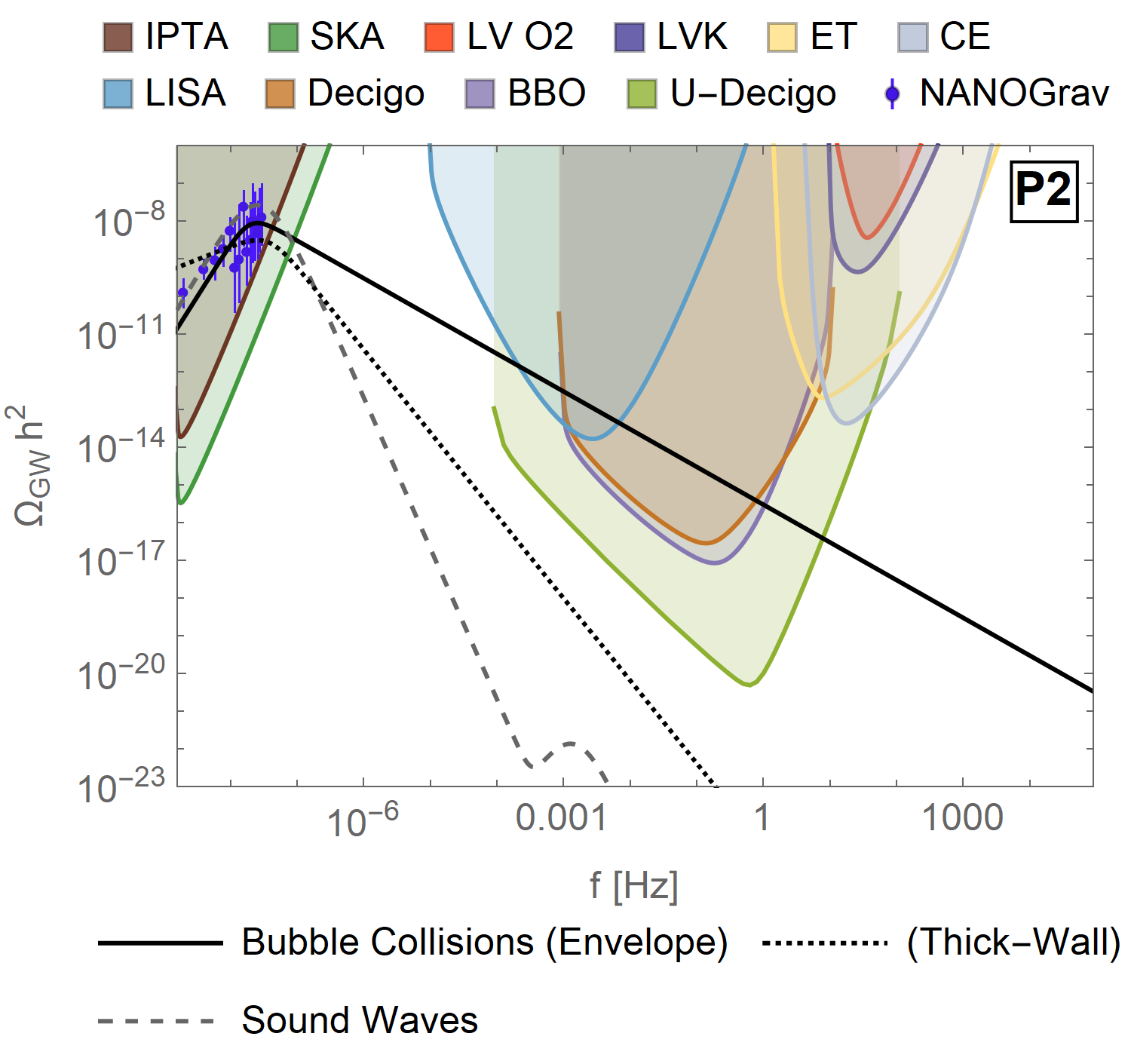}\\[1.mm]
\includegraphics[width=0.45\textwidth,height=0.40\textwidth]{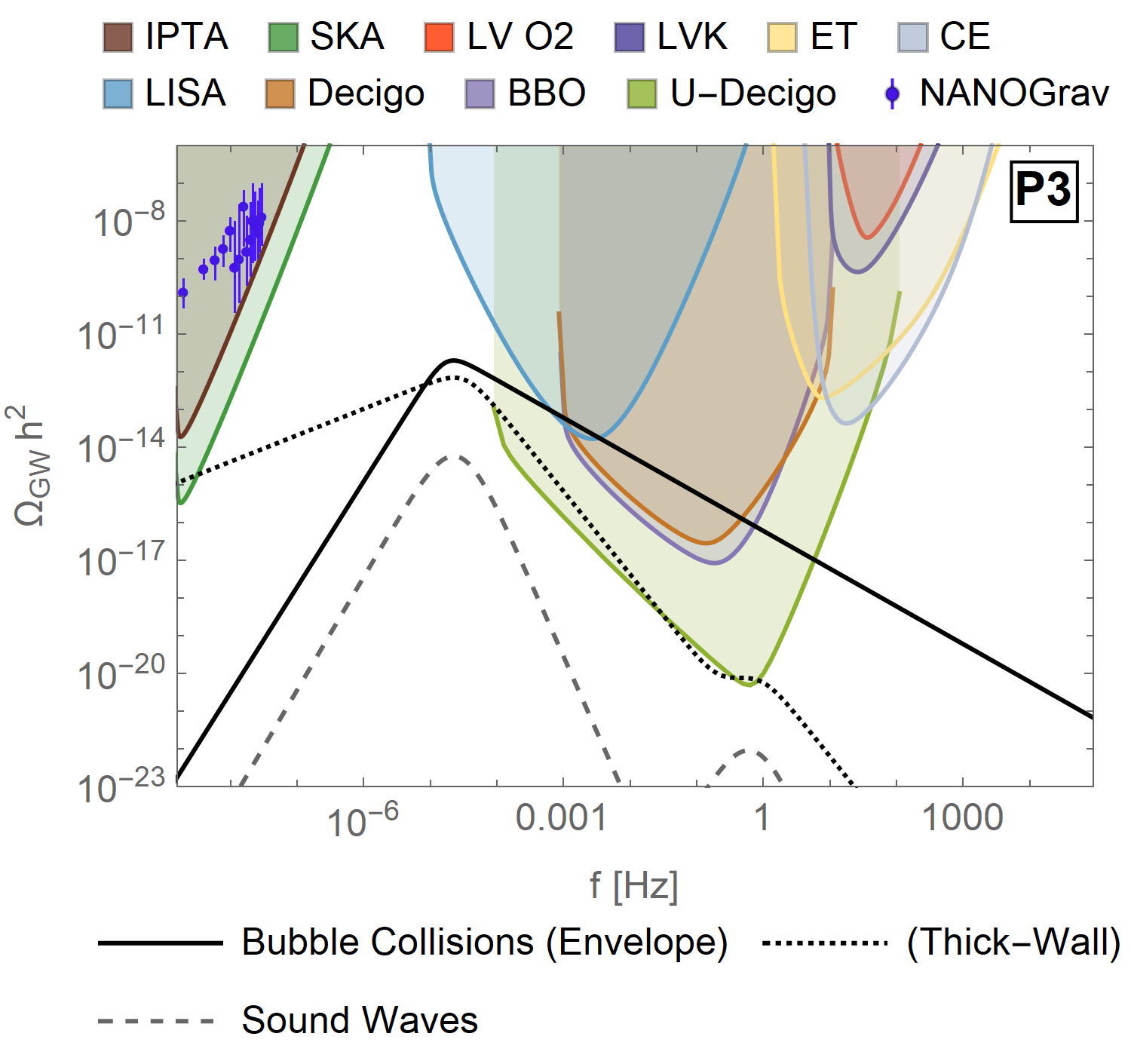}\hspace{2mm}
\includegraphics[width=0.45\textwidth,height=0.40\textwidth]{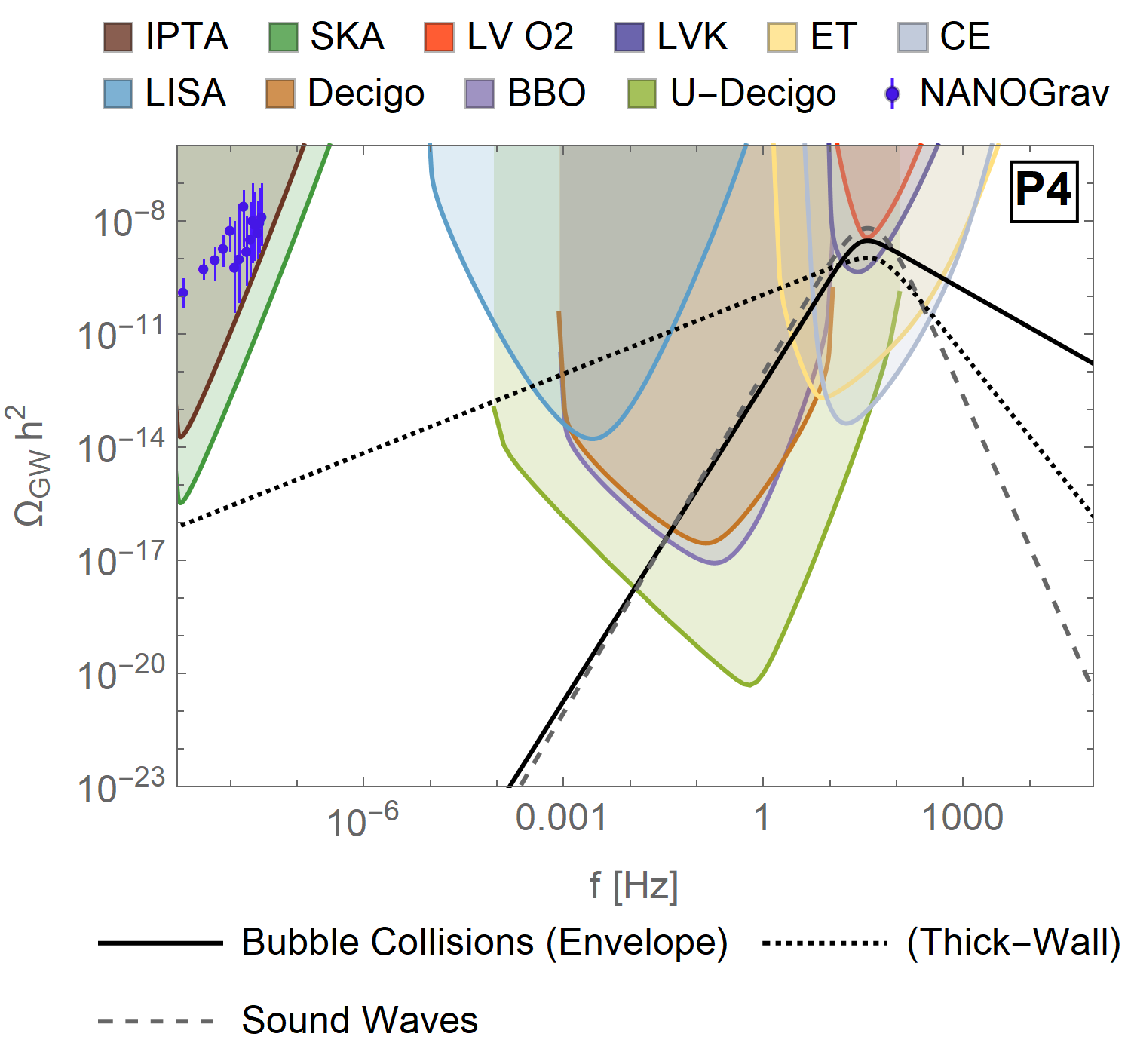}
\end{center}
\vspace{-6mm}
 \caption{
Parameter scans and gravitational wave spectra in the effective theory of chain inflation (case with a slower phase transition during the graceful exit). Upper panels, middle left panel: Parameter scans assuming that the gravitational wave signal is dominantly produced by bubble collisions (upper left: envelope approximation, upper right: thick-wall simulation) or by sound waves (middle left). In the blue-pink region chain inflation explains the NANOGrav signal. The other color-shaded regions indicate the parameter space testable by the experiments indicated in the plot legend (for acronyms see Sec.~\ref{sec:sensitivity_projections}). The striped regions are excluded by BBN and/or by the bubble percolation condition.
 Middle right panel and lower panels: gravitational wave spectra for the benchmark points P2, P3 and P4 (see Tab.~\ref{tab:Benchmark points}) indicated in the parameter scans. Also shown are the NANOGrav data and the projected sensitivities (power-law-integrated sensitivity curves) of various gravitational wave observatories. The benchmark point P2 can explain the NANOGrav data, while P4 can be tested at aLIGO-aVIRGO-KAGRA within the next years.}
\label{fig:postcritical}
\end{figure}

In order to explore the experimental prospects for detecting the gravitational waves associated with the graceful exit from chain, we performed parameter scans in the $\alpha-V_*^{1/4}$-plane (while we fixed $S_2=1$, $S_3=4$, $S_4=5$ for concreteness). The first three panels of Fig.~\ref{fig:postcritical} show the parameter regions which can be probed by present and future gravitational wave observatories. We again depict separately the cases in which the gravitational waves stem from (1) the bubble collision (modeled in the envelope approximation), (2) the bubble collision (modeled in the thick-wall simulation) (3) sound waves in the plasma (see Sec.~\ref{sec:originGW}). The striped regions are excluded by BBN and/or the violation of the percolation condition\footnote{In the case of $\alpha>1$ the Universe enters a second period of vacuum domination before the final phase transition. If this period lasted too long - which is the case for $\alpha>21$ -- one runs into the empty-universe problem of old inflation: the vacuum bubbles nucleated during the last phase transition never percolate and the phase transition never completes.~\cite{Freese:2022qrl}}. In all three cases, the largest part of the viable chain inflation parameter space is accessible to gravitational wave observatories-- implying very exciting prospects to detect the gravitational wave signal from the graceful exit.

Pulsar timing arrays are particular sensitive to inflation scales $V_*^{1/4}= 10\:\text{MeV}-1000\:\text{GeV}$ which feature a graceful-exit-induced gravitational wave spectrum peaking in the nHz$-$\textmu Hz regime. Space-interferometers can probe $V_*^{1/4}= (10^3-10^9)\:\text{GeV}$ for which the spectrum peaks at mHz$-$Hz, while ground interferometers will access even higher inflation scales $V_*^{1/4}= (10^9-10^{12})\:\text{GeV}$ with gravitational wave peak frequencies of Hz$-$kHz. The magnitude of the signal grows with $\alpha$ since larger $\alpha$ implies a later phase transition (i.e.\ a larger scale factor entering Eq.~\eqref{eq:gravityspectrum} and Eq.~\eqref{eq:gravityspectrumsound}). Parameter regions with $\alpha\gtrsim 0.1$ -- which feature a second period of vacuum-significance or vacuum-domination after inflation -- induce a gravitational wave signal which is already in reach for ongoing PTA or interferometer experiments.

\begin{table}[t]
\begin{center}
\begin{tabular}{|cccc|}
\hline
&&&\\[-4mm]
 & $\quad V_*^{1/4}\, \text{(GeV)}\quad$ & $\quad S_2 + S_3 + S_4\quad$ & $\quad \alpha \quad$ \\
 \hline\hline
&&&\\[-4mm] 
 P1 & $6.3\times 10^3$ & $19.6$ &  $0$ \\
 \hline
 P2 & $5.0$ & $10$ &  $0.7$\\
 \hline
 P3 & $1.8 \times 10^3$ & $10$ &  $0.01$\\
 \hline
 P4 & $10^{10}$ & $10$ &  $0.5$\\ \hline
\end{tabular}
\end{center}
\vspace{-0.4cm}
\caption{Benchmark points P1, P2, P3, P4 in Figures \ref{fig:nopostcritical} and \ref{fig:postcritical}. Here, $V_*$ corresponds to the scale of inflation at horizon crossing of the pivot scale. The expansion parameters $S_i$ and the strength of the phase transition $\alpha$ are defined in Eqs.~\eqref{eq:Sexpansion} and \eqref{eq:alpha}, respectively.}
\label{tab:Benchmark points}
\end{table}

A very intriguing possibility is that the NANOGrav signal for a stochastic gravitational wave background originates from the graceful exit from chain inflation. In Fig.~\ref{fig:postcritical} the NANOGrav favored region (2$\sigma$-region) is indicated in blue with the pink boundary. It can be seen that the size of the NANOGrav region depends on the gravitational wave emission mechanism (bubble collisions vs. sound waves) and the modeling of the spectrum (envelope approximation vs. thick-wall simulation). For instance, the envelope approximation predicts a harder infrared tail of the spectrum compared to the thick-wall simulation. As a consequence the peak frequency of the spectrum needs to fall into the NANOGrav sensitivity window in order to explain the signal for the envelope case (because the infrared tail has a too steep power-law to explain the data), while it can also be above the sensitivity window for the thick-wall simulation. This results in the NANOGrav-favored region extending to larger inflation scales $V_*^{1/4}\sim 10^3\:\text{GeV}$ for the thick-wall simulation compared to the envelope approximation. But independent of the modeling of the gravitational wave spectrum, we always find chain inflation parameter space which can explain NANOGrav. For instance, in the middle right panel of Fig.~\ref{fig:postcritical} we depict the gravitational wave spectrum for the benchmark point P2 (with $\alpha=0.7$, $V_*^{1/4}=5.0\:\text{GeV}$) which crosses right through the NANOGrav data (for all three considered emission cases). Intriguingly, the graceful exit phase transition provides a significantly better fit to the NANOGrav data compared to the standard astrophysical explanation in terms of black hole mergers (the latter has been found to be just barely consistent with NANOGrav at the $2\sigma$-level~\cite{NANOGrav:2023hvm}).

Even if the NANOGrav signal turns out to have a different (e.g.\ astrophysical) origin there still exist great prospects for a chain inflation discovery in the very near future. The benchmark point P4 (with $\alpha=0.5$, $V_*^{1/4}=10^{10}\:\text{GeV}$), for example, induces a gravitational wave signal which can be tested at the aLIGO-aVIRGO-KAGRA network within the next years. The corresponding gravitational wave spectrum is depicted in the lower right panel of Fig.~\ref{fig:postcritical}.

Clearly, for any present (NANOGrav) or future gravitation wave signal, we want to discriminate chain inflation against other possible sources. A unique feature that we can explore is related to the superposition of waves from during chain inflation and from a potentially extended graceful exit. Because a graceful exit phase transition is much slower than the phase transitions during inflation, it produces gravitational waves with a substantially lower peak frequency. We thus expect a double-peak spectrum with a low-frequency peak from the extended graceful exit and a fainter high-frequency peak from the inflationary epoch. For large $\alpha\gtrsim 0.1$, the first peak (from the graceful exit) is so much stronger that it completely overshadows the second peak. This is the case for the benchmark points P2 and P4 in Fig.~\ref{fig:postcritical}. If $\alpha<0.1$, on the other hand, the second peak (from during inflation) can potentially be tested. The lower left panel of Fig.~\ref{fig:postcritical} depicts such a double-peak spectrum which is obtained for the benchmark point P3 (with $\alpha=0.01$, $V_*^{1/4}=1.8\:\text{TeV}$).\footnote{The double-peak spectrum occurs for gravitational waves from sound waves. It also occurs for gravitational waves from bubble collisions if we model the spectrum according to the thick-wall simulation. On the other hand, the second peak does not occur if we employ the envelope approximation. This is because of the softer ultraviolet tail of the graceful-exit-induced spectrum in the envelope approximation which overlays the inflation-induced spectrum. We note, however, that the soft ultraviolet tail is linked to the unphysical assumption of vanishing shear stress immediately after the bubble collisions which is made in the envelope approximation. Making the (more realistic) assumption of propagating shear stress as in the thick-wall simulation, the gravitational wave spectrum is expected to decrease more rapidly above the first peak. Hence, we consider the occurrence of a double-peak spectrum (for $\alpha\lesssim 0.1$) a physical reality.} Future space interferometers can potentially establish the double-peak structure and provide strong evidence for chain inflation as a source of the gravitational wave signal.

Let us emphasize, however, that even if only a single-peak gravitational wave spectrum is measured, this would already hint at a phase-transition origin because astrophysical sources do typically not produce sharply peaked spectra. In addition, complementary cosmological observables -- for instance related to the scale-dependence of the scalar power spectrum (see Sec.~\ref{sec:scaledependence}) -- can be employed to single out chain inflation as the source of a gravitational wave signal.

\section{Summary and Conclusion}

In chain inflation the expanding Universe undergoes a series of first-order phase transitions between different vacua. While the emerging picture is markedly different from slow-roll inflation it is consistent with all cosmological constraints. In particular, chain inflation produces a nearly scale-invariant scalar power spectrum of density perturbations as observed by CMB experiments. But in contrast to slow-roll inflation, the density perturbations do not result from quantum fluctuations of the inflaton. Rather, they are linked to the probabilistic nature of vacuum tunneling which implies that different patches of the Universe undergo the vacuum transitions at slightly different times.

In this work we derived the gravitational wave spectrum of chain inflation. While in slow-roll inflation gravitational waves predominantly originate from the primordial tensor fluctuations of the metric, such metric fluctuations are suppressed in chain inflation due to an upper bound $V_*^{1/4}\lesssim 10^{12}\:\text{GeV}$ on the scale of inflation. Nevertheless, we find that chain inflation gives rise to spectacular gravitational wave signatures. The latter are sourced by the bubble collisions and the sound waves created in the radiation plasma during the first-order phase transitions. As a consequence of the different origin, the frequency range of the produced gravitational waves is very different in chain inflation compared to slow-roll inflation. Gravitational waves from slow-roll inflation in the aHz$-$fHz-regime can in optimistic cases be detected indirectly through a polarization signal in the CMB. In contrast, the gravitational wave spectrum from chain inflation peaks in the nHz-kHz-regime -- implying exciting discovery potential for direct gravitational wave searches with pulsar timing arrays and interferometer experiments.

 The gravitational wave signal from chain inflation generically has two components: a faint broad peak from the collection of phase transitions during inflation and a more intense (lower-frequency) peak from a potentially extended graceful exit from inflation. During chain inflation, the Universe populates very short-lived vacua, but today it has settled in a (meta)stable vacuum. In between, there likely occurs a vacuum with an intermediate lifetime whose decay marks the graceful exit from inflation. Since the gravitational wave amplitude is proportional to the duration of the phase transition, this extended graceful exit produces a particular strong gravitational wave signal.  Even if only a single-peak gravitational wave spectrum is measured, this would already hint at a phase-transition origin because astrophysical sources do typically not produce sharply peaked spectra. 

We systematically investigated the sensitivity of gravitational wave experiments to the signature from chain inflation. Intriguingly, there exist great prospects to detect the gravitational waves from the graceful exit already in the very near future (see Fig.~\ref{fig:postcritical}). Chain inflation at a scale $V_*^{1/4}= (10^9-10^{12})\:\text{GeV}$ produces a signal which could be just around the corner for ongoing ground-interferometer experiments like aLIGO and aVIRGO, while future space missions like LISA and BBO are particular sensitive to inflation scales $V_*^{1/4}= (10^3-10^{9})\:\text{GeV}$. Finally, low-scale chain inflation with $V_*^{1/4}=10\:\text{MeV}-1000\:\text{GeV}$ can be tested at pulsar timing array experiments. A particularly exciting possibility is that the graceful exit from chain inflation is responsible for the stochastic gravitational wave background (tentatively) discovered at the NANOGrav PTA. We performed a detailed statistical analysis of the latest NANOGrav data set and, indeed, found the signal to be compatible with chain inflation at a scale $V_*^{1/4}=(1-1000)\:\text{GeV}$. Such low inflation scales are rarely considered in the literature because, in slow-roll inflation, they would cause a severe fine-tuning problem (by imposing an insanely flat inflaton potential to obtain the correct CMB normalization). However, due to the different origin of CMB anisotropies, low-scale chain inflation -- as required to explain the NANOGrav signal -- does naturally arise without any fine-tuning of the potential. Chain inflation, in fact, provides a significantly better fit to the NANOGrav gravitational wave spectrum compared to astrophysical explanations in terms of black hole mergers.

For any present (NANOGrav) or future gravitational wave discovery, it is crucial to distinguish chain inflation against other possible sources. A very characteristic feature of chain inflation is the mentioned double-peak structure of the gravitational wave spectrum. While the prospects of detecting the stronger peak from the graceful exit are particularly bright, there is also the possibility to measure the second fainter, broader peak from the phase transitions during inflation. We found that advanced future space interferometers like BBO and U-Decigo will potentially be able to detect the second peak. If a double-peak gravitational wave spectrum can indeed be established, this would amount to a dramatic confirmation of the prediction from chain inflation.

But even if only a single-peak gravitational wave spectrum is found, there exist complementary cosmological probes which could link the signal to chain inflation. In particular, chain inflation tends to produce a
deviation of the scalar power spectrum from the power-law form which is larger than in slow-roll inflation. We identified significant chain inflation parameter space, in which this signature can be detected at CMB experiments such as the Simons Observatory (taking data very soon) and the proposed PIXIE satellite.

\section*{Acknowledgments}

K.F.\ is Jeff \& Gail Kodosky Endowed Chair in Physics at the University of Texas at Austin, and K.F.\ and M.W.\ are grateful for support via this Chair. K.F.\ and M.W.\ acknowledge support by the U.S. Department of Energy, Office of Science, Office of High Energy Physics program under Award Number DE-SC-0022021. K.F., A.L.\ and M.W.\ acknowledge support by the Swedish Research Council (Contract No. 638-2013-8993).

\bibliography{chainbib}
\bibliographystyle{h-physrev}

\end{document}